\documentclass[reprint,twocolumn,secnumarabic,amssymb, aps, prx]{revtex4-2}

\usepackage{graphicx}
\usepackage{gensymb}
\usepackage{amsmath}
\usepackage{braket}
\usepackage{lipsum}
\usepackage[colorlinks,linkcolor=black,citecolor=black,urlcolor=black,filecolor=black]{hyperref}
\usepackage{wrapfig}
\usepackage{listings}
\lstdefinestyle{mystyle}{
    basicstyle=\ttfamily\footnotesize,
    breakatwhitespace=false,         
    breaklines=true,                 
    captionpos=b,                    
    keepspaces=true,                 
    numbers=left,                    
    numbersep=5pt,                  
    showspaces=false,                
    showstringspaces=false,
    showtabs=false,                  
    tabsize=2
}

\begin{document}
\title{Architectural mechanisms of a universal fault-tolerant quantum computer}

\author{
Dolev~Bluvstein$^{1,*}$, Alexandra~A.~Geim$^{1,*}$, Sophie~H.~Li$^{1}$, Simon~J.~Evered$^{1}$, J. Pablo Bonilla Ataides$^{1}$, Gefen~Baranes$^{1,2}$, Andi~Gu$^{1}$, Tom~Manovitz$^{1}$, Muqing~Xu$^{1}$, Marcin~Kalinowski$^{1}$, Shayan~Majidy$^{1}$, Christian~Kokail$^{3,1}$, Nishad~Maskara$^{1}$, 
Elias~C.~Trapp$^{1}$, 
Luke~M.~Stewart$^{1}$, 
Simon~Hollerith$^{1}$, Hengyun~Zhou$^{1,4}$,  Michael~J.~Gullans$^{5}$, Susanne~F.~Yelin$^{1}$, Markus~Greiner$^{1}$, Vladan~Vuleti\'{c}$^{2}$, Madelyn~Cain$^{1}$, and Mikhail~D.~Lukin$^{1}$}
\affiliation{$^1$Department~of~Physics,~Harvard~University,~Cambridge,~MA~02138,~USA \quad \quad \\ 
$^2$Department~of~Physics~and~Research~Laboratory~of~Electronics,~Massachusetts~Institute~of~Technology,~Cambridge,~MA~02139,~USA\\
$^3$ITAMP, Harvard-Smithsonian Center for Astrophysics, Cambridge, MA 02138, USA\\
$^4$QuEra Computing Inc., Boston, MA 02135, USA\\
$^5$Joint Center for Quantum Information and Computer Science, NIST/University of Maryland, College Park, Maryland 20742, USA\\
$*$ These authors contributed equally to this work.
}

\begin{abstract}

Quantum error correction (QEC) \cite{Shor1996, Steane1996} is believed to be essential for the realization of large-scale quantum computers \cite{Benioff1980, Preskill2018}. However, due to the complexity of operating on the encoded `logical' qubits \cite{Aharonov1999,Dennis2002a}, understanding the physical principles for building fault-tolerant quantum devices and combining them into efficient architectures is an outstanding scientific challenge. Here we utilize reconfigurable arrays of up to 448 neutral atoms to implement all key elements of a universal, fault-tolerant quantum processing architecture and experimentally explore their underlying working mechanisms. We first employ surface codes to study how repeated QEC suppresses errors \cite{Dennis2002a, Acharya2024}, demonstrating $2.14(13)$x below-threshold performance in a four-round characterization circuit by leveraging atom loss detection and machine learning decoding \cite{Wu2022, Baranes2025}. We then investigate logical entanglement using transversal gates and lattice surgery \cite{Horsman2012, Bluvstein2023,Cain2025}, and extend it to universal logic through transversal teleportation with 3D [[15,1,3]] codes \cite{Raussendorf2012, Bombin2015}, enabling arbitrary-angle synthesis with logarithmic overhead \cite{Aharonov1999,Dawson2005}. Finally, we develop mid-circuit qubit re-use  \cite{Wu2019}, increasing experimental cycle rates by two orders of magnitude and enabling deep-circuit protocols with dozens of logical qubits and hundreds of logical teleportations \cite{Gottesman1999, Knill2005, Raussendorf2001, Sahay2023a} with [[7,1,3]] and high-rate [[16,6,4]] codes while maintaining constant internal entropy. Our experiments reveal key principles for efficient architecture design, involving the interplay between quantum logic \& entropy removal, judiciously using physical entanglement in logic gates \& magic state generation, and leveraging teleportations for universality \& physical qubit reset. These results establish foundations for scalable, universal error-corrected processing and its practical implementation with neutral atom systems.

\end{abstract}
\maketitle

The central challenge of quantum computation is its inherent sensitivity to errors \cite{Preskill2018,Landauer1995}. Whereas classical computers are composed of intrinsically robust digital bits stabilized by dissipation \cite{Landauer1961}, quantum states are intrinsically analog objects that evolve in a continuous state space, with coherent unitary evolution that does not allow dissipation \cite{Landauer1995}. Remarkably, the discovery of quantum error correction (QEC) \cite{Shor1995,Steane1996,Kitaev1997} provides a method to realize robust quantum computation. In this approach, entanglement between many physical qubits is used to encode error-corrected `logical' qubits that can have exponentially low, digitized errors while performing arbitrary, analog-like computations \cite{Shor1996,Aharonov1999}.

While recent experiments provide clear examples where QEC works in practice \cite{Bluvstein2023,Acharya2024,Reichardt2024a}, large-scale fault-tolerant quantum computation (FTQC) remains a formidable challenge. In essence, it requires one to dissipatively remove errors from the physical system, while simultaneously performing arbitrary coherent manipulation of the encoded logical qubits. Such operation requires a broad range of components \cite{Reichardt2024,SalesRodriguez2024}, coupled with many important hardware considerations and a wide range of QEC techniques \cite{Sivak2023, Bravyi2024,Cain2025,Gidney2024,Putterman2025}. Implementing and combining these techniques in practical systems, and understanding the scientific principles of resulting fault-tolerant architectures, is a highly complex task at the frontier of quantum science.

\begin{figure*}
 \includegraphics[width=2\columnwidth]{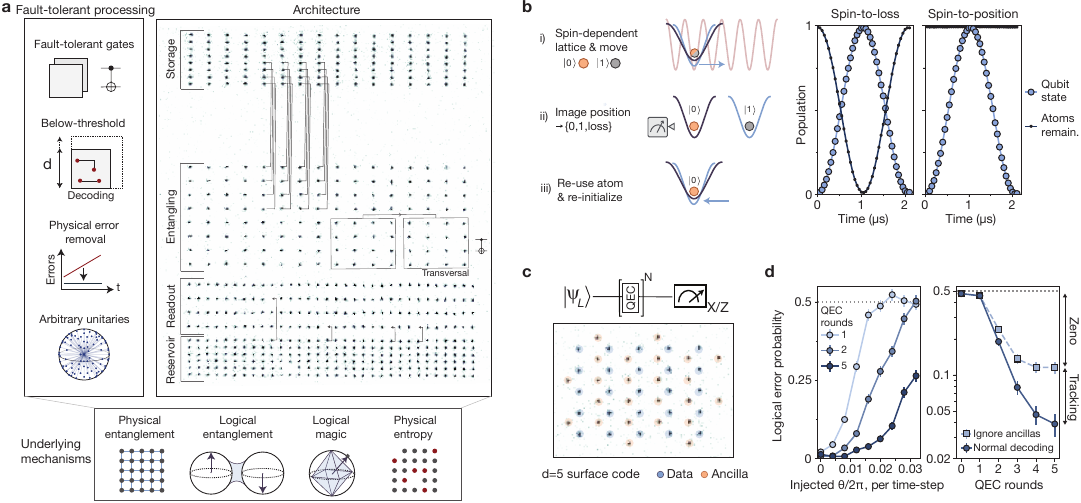}
\caption{\textbf{Architectures and mechanisms for fault-tolerant quantum computation.} \textbf{a,} We study the key building blocks of fault-tolerant processing. We utilize an architecture based on reconfigurable atom arrays trapped in optical tweezers, where the logical processor is segmented into storage, entangling, readout, and reservoir zones. Underlying physical mechanisms are identified and characterized. \textbf{b,} Spin-to-position conversion for non-destructive, loss-resolved qubit readout is accomplished with a state-selective 1D lattice that converts the atom spin state into position. Plot shows measured Rabi oscillation. This non-destructive readout has 0.46(4)\% bit-flip error and 0.24(2)\% loss (Methods). \textbf{c,d,} Stabilizer measurement on a $d=5$ surface code is interspersed with global coherent errors injected on the data qubits. Each CZ layer corresponds to one time-step (Methods). Repeated correction reduces error build-up through both the Zeno effect and error tracking (right plot is at a fixed $\theta/2\pi = 0.016)$. For visual clarity, an acceptance fraction of 50\% is used in this plot (Methods).
}
\label{fig1}
\end{figure*}

Here we experimentally implement the core building blocks for scalable, universal quantum computation and identify and explore key mechanisms for efficient architectures suitable for deep-circuit FTQC. Utilizing a quantum processor based on reconfigurable neutral atom arrays and leveraging key hardware upgrades, we study below-threshold error correction, fault-tolerant quantum operations, arbitrary unitary synthesis, and physical error removal during deep-circuit execution, altogether achieving all ingredients for scalable (i.e., logarithmic overhead) realization of arbitrary quantum algorithms \cite{Aharonov1999}. Our results further advance the quantum computing frontier along several directions: (i) advanced decoding methods leveraging erasure information \cite{Wu2022, Baranes2025} and machine learning \cite{Bausch2024} are used to optimize below-threshold performance; (ii) syndrome measurements and physical entanglement are reduced to being used only where they are needed during universal logic; (iii) teleportation \cite{Gottesman1999, Knill2005, Raussendorf2001, Sahay2023a} is used to achieve efficient universality with transversal operations, and enable physical qubit reset without additional overhead in the algorithm. These result in significant improvements to architecture design and orders-of-magnitude reduction to space and time overheads.

\subsection*{Neutral atom logical processor}

To implement these core building blocks, our experiments utilize a logical processor \cite{Bluvstein2023} with up to 448 atoms. Qubits are stored in the hyperfine clock states of $^{87}$Rb atoms trapped in optical tweezers generated by a spatial light modulator (SLM) in storage, entangling, readout, and reservoir zones (Fig.~1a). Quantum circuits are programmed by shuttling qubits in the middle of the computation with a 2D acousto-optic deflector \cite{Beugnon2007, Schlosser2011, Bluvstein2022}, high-fidelity entangling operations are realized via fast excitation to Rydberg states \cite{Evered2023a}, and fully programmable single-qubit operations are realized via locally focused Raman beams \cite{Bluvstein2023}. Crucially, this system provides control parallelism over logical qubit blocks \cite{Bluvstein2023}: for logic gates, repeated error correction, and even universal computation, all the physical qubits within the logical block realize identical operations with parallel instructions delivered by optical controls.

\begin{figure*}
\includegraphics[width=2\columnwidth]{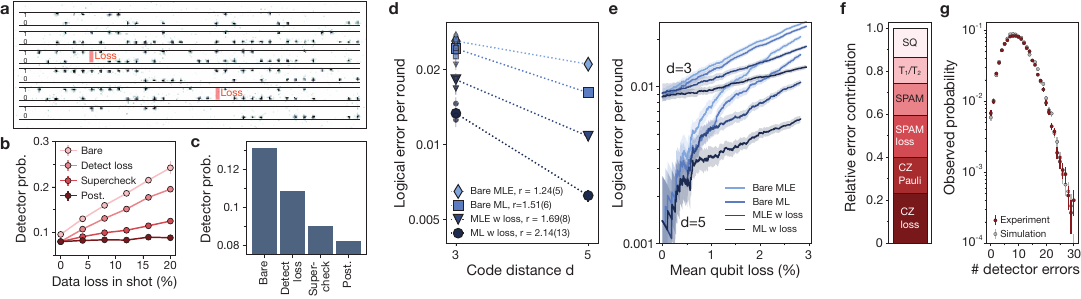}
\caption{\textbf{Below-threshold repeated quantum error correction leveraging loss detection.} \textbf{a,} Results of repeated rounds of $d=5$ surface code using loss detection, showing a snapshot of the data block and multiple ancilla blocks (see ED Fig.~\ref{fig:ED_ExtraSurface}). \textbf{b, c,} Products of stabilizer measurement results between rounds are used to detect qubit errors, which we refer to as `detectors'. `Bare' counts loss as state $\ket{0}$, `detect loss' does not make this erroneous assignment, `supercheck' multiplies detectors around lost atoms, and `post.' postselects on all atoms of a detector being present. \textbf{b,} Detector error probability as a function of data qubit loss in each shot, analyzed by partitioning the total dataset. \textbf{c,} Average over all data. \textbf{d,} Logical error per round for a surface code after 4 QEC rounds in both bases, decoded using most-likely error methods (`bare MLE'), machine learning (`bare ML'), MLE with loss information, and ML with loss information. ML with loss renders the error per round for $d=5$ as $2.14(13)$x lower than $d=3$. No postselection is used. Small points are the four d=3 quadrants. Results are averaged between $\ket{+_L}$ and $\ket{0_L}$ initialization bases. See Methods and ED Fig.~\ref{fig:ED_ExtraSurface} for more details.
\textbf{e,} Logical error per round as a function of the mean qubit loss, plotted as a cumulative density function. \textbf{f,} Relative physical error contribution to overall error budget (see Methods). \textbf{g,} Distribution of detector errors per shot, suggesting the absence of large-scale correlated errors.
}
\label{fig2}
\end{figure*}

A key experimental upgrade involves non-destructive, spin-resolved qubit readout using a one-dimensional optical lattice in the readout zone (Fig.~1b). Whereas conventional readout is realized via spin-to-loss conversion followed by camera readout \cite{Bluvstein2023}, the state-selective lattice enables splitting the two qubit states into two separate tweezers \cite{Wu2019}, thereby realizing spin-to-position conversion that enables both loss detection and atom retention. Combined with techniques for mid-circuit re-initialization, this enables qubit re-use for extended computation as well as a two orders of magnitude increase to the experimental cycle rate, as described below.  

Another key addition involves the use of repeated error correction for removing errors produced during quantum algorithms. To probe this function of QEC, we use a distance-5 surface code logical qubit \cite{Dennis2002a,Acharya2024} with up to five rounds of stabilizer measurement (Figs.~1c,d). Here, data qubits are held in static potentials in the entangling zone, and an ancilla block is moved over to perform parallel entangling operations required for measurements of stabilizers. After performing  the stabilizer checks, the ancilla block is moved to the storage zone, and a new block is brought in to repeat the cycle. Upon injecting global coherent errors at regular intervals, we find repeated correction significantly suppresses logical errors, as coherent errors are projected into incoherent ones (with quadratic suppression) due to the quantum Zeno effect \cite{Itano1990} and tracked. Notably, although the coherent errors are injected as global, correlated single-qubit rotations, stabilizer measurements convert these into uncorrelated, incoherent physical errors \cite{Bravyi2018New}, and prevent coherent errors on the logical qubit level (Methods). Fig.~1d highlights the function of repeated QEC: the conversion of general errors to correctable bit-flip and phase-flip errors removes entropy by reducing the range of possible errors \cite{Aharonov1999}, and detecting the incoherent errors and using them in decoding reduces entropy by keeping track of the state of the system \cite{Dennis2002a}. 

\subsection*{Below-threshold performance}

We first utilize these tools to explore below-threshold performance of a surface code qubit under multiple rounds of QEC. Theoretically, logical errors can be exponentially suppressed as $(p/p_{\text{th}})^{(d+1)/2}$ by increasing code distance $d$ if the physical error rate $p$ is below a characteristic threshold $p_{\text{th}}$ \cite{Aharonov1999,Dennis2002,Acharya2024}. Figure~2a shows the measurement results of the data qubits and multiple sets of ancilla qubits in the configuration shown in Fig.~1c, obtained with non-destructive, spin-resolved readout (Fig.~1b), which identifies which atoms were lost during the quantum circuit. We observe that qubit loss events result in flickering stabilizer patterns \cite{Stace2009} (see ED Fig.~\ref{fig:ED_SurfaceLoss}b and Methods) which create time-correlations between detected error events and leads to a sharp rise in detected errors (Fig.~2b). 

To correct for the effect of atom loss, we construct so-called superchecks (see Ref.~\cite{Stace2009} and Methods) by multiplying stabilizers around the lost atom to recover error information, which results in a suppressed detected error probability (Fig.~2b). These observations indicate that using the loss information in decoding can greatly improve QEC performance, as predicted theoretically \cite{Wu2022, Baranes2025}. To take advantage of these features, we use most-likely error (MLE) \cite{Cain2024a,Baranes2025} and machine learning decoders \cite{Bausch2024} trained on simulated and experimental data, incorporating loss information into both (Methods).

\begin{figure*}
\includegraphics[width=2\columnwidth]{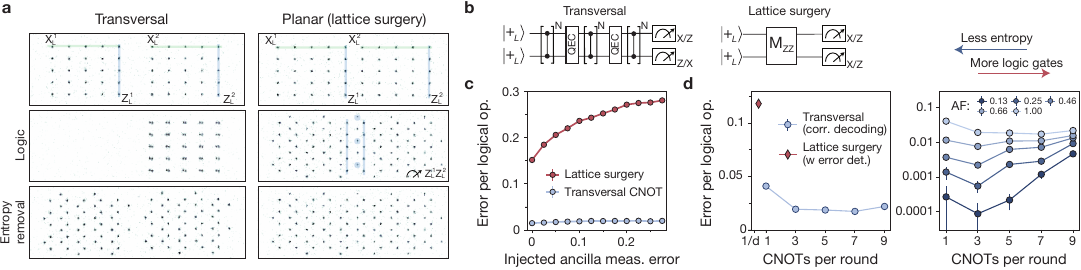}
\caption{\textbf{Exploring the interplay of logic gates and entropy removal.} 
\textbf{a,} Atom images illustrating two-qubit logic gates and stabilizer measurement. Lattice surgery is realized using ancillas to measure the logical product $Z^1_L Z^2_L$, and transversal gates are realized via atomic motion. \textbf{b,} Quantum circuits for realizing  transversal gates and lattice surgery operations. CZ gates are realized as transversal CNOTs and Hadamards. \textbf{c, } Dependence of error of the logic operation on the ancilla measurement error. Lattice surgery errors rapidly worsen with increasing ancilla measurement errors (injected in post-processing). \textbf{d, } N repeated logic operations are interspersed with rounds of QEC stabilizer measurement. Transversal CNOT has a lower error than lattice surgery (left), and has an optimum of roughly 3 CNOTs per round, as seen most clearly when modest postselection, characterized by  acceptance fractions (AF) are used (right). Error detection is used for lattice surgery in \textbf{d} to compensate for having $<d$ rounds (Methods).
}
\label{fig3}
\end{figure*}

We now characterize repeated QEC performance as a function of the code distance $d$ \cite{Bluvstein2025a}. We find that $d$=5 has a 2.14(13)x lower error per round than $d$=3, indicating below-threshold behavior for this four-round characterization circuit (Fig.~2d). In particular, we find that the loss information and machine learning decoding together improve the QEC performance by a factor of 1.73(13)x compared to conventional methods. Note that while our observations show below-threshold behavior for repeated QEC rounds, theoretically the threshold can get worse by $\approx$ 1.15x in the limit of many repeated rounds and by $\approx$ 1.1x when incorporating $\approx 1$ transversal gate per QEC round (see Methods). The error per round is 0.62(3)\% for $d=5$, and reduces toward 0.1\% per round on shots with no qubit loss (Fig.~2e), consistent with a $p^3$ scaling and roughly half of our errors being loss.

These observations are consistent with numerical simulations using a simple empirical error model based on separately characterized error rates (see error budget in Fig.~2f and Supplementary Information). We further note that our observed distribution of errors-per-shot is closely consistent with these simple simulations with uncorrelated Pauli-type and loss-type errors, suggesting the absence of large-scale correlated errors in the system. We observe that time-correlations are almost fully diminished when postselecting on no qubit loss, indicating that almost all leakage ($>80\%$, Methods) corresponds to atom loss, while leakage to other hyperfine states appears to be suppressed. This observation suggests that various scattering channels \cite{Cong2022} naturally loss-convert, likely due to anti-trapping of metastable states.

\subsection*{Stabilizer measurement during logic operations}

We next extend repeated QEC to logical operations, by studying two different approaches for realizing quantum logic (Fig.~3a), where the stabilizer measurements play different roles. In the transversal method, a logical gate is realized by physically transporting and interlacing the data qubits of two surface code blocks, and then applying pairwise entangling gates \cite{Shor1996, Dennis2002a, Bluvstein2023}. In the planar setting, logical entanglement is realized via lattice surgery \cite{Horsman2012}, where a joint logical measurement is performed by ancilla qubits added between the two codes (Figs.~3a,b). To probe these logical operations, in the transversal approach, we apply three rounds of $N$ repeated CNOTs interleaved with two rounds of stabilizer checks.  In the lattice surgery approach, we carry out a $Z^1_L Z^2_L$ joint logical measurement using two rounds of stabilizer measurements \cite{Horsman2012}. We then measure the $X^1_L X^2_L$ and $Z^1_L Z^2_L$ parities of the resulting logical Bell state for both methods. 

To test the role of stabilizer measurements in both cases, we first introduce additional ancilla measurement errors in postprocessing (Fig.~3c). We observe a lower logical error for transversal gates, and that the lattice surgery is significantly more sensitive to injected measurement errors. To explore the interplay of logic gates and entropy removal, we next study the transversal gate performance as a function of the number of repeated CNOT gates per QEC round. We find optimal performance for several CNOTs per QEC round, and using postselection on decoding confidence (see Methods), observe that approximately three CNOTs per round is optimal at low error rates (Fig.~3d). 

These results highlight multiple key aspects of FTQC. First, the response to ancilla measurement errors highlights the key distinction between the two approaches: in the transversal gate setting, the logic is realized directly between the data qubits which store the underlying logical states, and the role of stabilizer measurements is to remove entropy, whereas in lattice surgery, the ancilla measurements directly perform the logic operation and have to be correct. The need for correct stabilizer measurements is the origin of the conventional fault-tolerance assumption of $d$ rounds per QEC cycle \cite{Dennis2002a, Horsman2012}. Conversely, with transversal operations, the observed optimum at 3 CNOTs per round corresponds to stabilizer measurements balancing the local entropy generated by the logic gate. This is consistent with theoretical predictions for universal computation with correlated decoding techniques \cite{Cain2024a,Zhou2024,Cain2025}. Second, Fig.~3d clearly shows that the error per logical gate depends on the number of gates applied per QEC round. This highlights a key distinction between operations on physical and logical qubits: while physical gate performance can be well-characterized by a single fidelity value $F$, logical gate performance depends both on the decoding success probability - which reflects the physical entropy of the system, captured by the detector error probability $p_{\text{det}}$ - and on how much that entropy increases per logical gate, quantified by $\Delta p_{\text{det}}$. We find that a simple model for logical gate fidelity $F_L$, where $1-F_L \propto \left[(p_{\text{det}} + N \Delta p_{\text{det}})^{(d+1)/2} \right]/ N$, with the measured $\Delta p_{\text{det}}$ consistent with our physical two-qubit gate error, accurately describes the experimental data (Methods).

\begin{figure*}
\includegraphics[width=2\columnwidth]{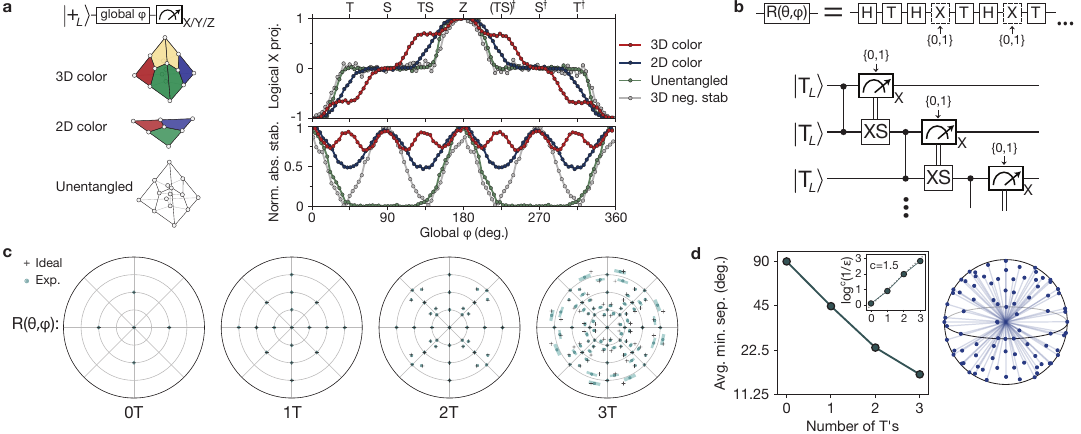}
\caption{\textbf{Synthesizing arbitrary-angle rotations with a universal fault-tolerant gate set.} \textbf{a,} Subjecting codes of various dimensionality to global rotations. Plateaus are seen at robust angles where the stabilizers also revive. 3D Reed-Muller codes, only with all positive stabilizer signs, have an additional plateau at 45-degrees corresponding to non-Clifford $T$ gates. \textbf{b,} Programmable angles $R(\theta, \phi)$ are realized by an alternating sequence of $H$ and $T$ gates. Circuit shows an implementation of such a sequence with $T$ implemented via state preparation and $H$ implemented via quantum teleportation. \textbf{c,} Polar angle plots show generated angles using entangled Reed-Muller codes, measured by tomography, for different maximum number of $T$ gates. Experimental results are consistent with theory within statistical error. \textbf{d,} Experimental results show that the minimum separation between generated angles decreases exponentially with length of sequence. Inset shows a rescaling where $\epsilon$ is the angular separation to the target angle. Bloch sphere shows all measured angles. For visual clarity, variable degrees of acceptance fraction are used (Methods). } 
\label{fig4}
\end{figure*}

\subsection*{Universality and synthesizing arbitrary unitaries}

We now study how to realize arbitrary error-corrected unitary operations. The core ingredient is 
the Solovay-Kitaev theorem, which  states that arbitrary single-qubit rotations can be approximated to exponential precision using only digital gates such as Hadamard $H$ and $T = e^{-i \frac{\pi}{8} Z}$ (a 45-degree rotation around the Z-axis) \cite{Kitaev1997, Aharonov1999, Dawson2005}. We note that while the Eastin-Knill theorem prohibits realizing such a universal gate set with unitary transversal operations \cite{Eastin2009}, it can be circumvented by the introduction of logical measurement which breaks the unitarity constraint \cite{Raussendorf2012}.

In our approach, non-Clifford $T$ gates and universal rotations are realized with efficient transversal circuits built out of teleportation with 3D codes. Figure~4a shows experimental measurements where we use quantum circuits (see Methods) to create 2D color codes (Steane codes) \cite{Steane1996}, and 3D color codes (Reed-Muller codes), and subject them to global phase rotations $\varphi$ around the Z-axis \cite{Raussendorf2012}. For comparison, we also show the results of similar measurements for unentangled physical qubits analyzed as 3D codes, and 3D codes with incorrect values of stabilizers. The various configurations show plateaus in the logical expectation value, and revivals in the stabilizers, for multiples of 90-degree angles (multiples of 180-degree for unentangled qubits). Crucially, 3D codes, prepared in the proper entangled states, also exhibit robustness at multiples of 45-degrees, corresponding to their transversal $T $ gate \cite{Bombin2015}.

To implement unitary synthesis by the circuit in Fig.~4b, transversal teleportation is utilized. Specifically, we create multiple Reed-Muller codes in the $\ket{T_L}$ state, and by applying logical CZ gates followed by X-basis measurements (with feedforwards applied in-software here), the logical information is teleported with an $H$ gate \cite{Raussendorf2012}. Figure~4c shows the set of angles generated by the circuit in Fig.~4b using up to three $T$ gates. The resulting logical states span a range of points on the Bloch sphere and match the expected angles with high precision (consistent with statistical uncertainty). We observe that the angular spacing between accessible states shrinks exponentially with the number of $T$ gates. This enables precise synthesis of rotation angles with a logarithmic number of steps (Fig.~4d inset), as expected by the Solovay-Kitaev theorem \cite{Kitaev1997, Aaronson2004, Aharonov1999}.

These observations demonstrate that teleportation can be used as a powerful tool for universal processing, allowing precise analog rotations to be built from digital gates. Interestingly, although the circuit realized here is fully transversal, the measurement, decoding, and logical feedforward ensures that the logical information propagates unitarily while physical-level dissipation serves to correct errors. ED Fig.~\ref{fig:ED_Universality} explores code-switching and transversal gates between 2D and 3D codes, again finding the underlying source of universality is the measurement of the 3D code. In addition, we note that the role of stabilizers differs fundamentally when generating logical magic. While for Clifford circuits, stabilizer measurements are used to remove entropy, for transversal $T$ gates, correct stabilizer signs (e.g., $+1$ eigenvalues) are essential for realizing non-Clifford operations (Fig.~4a). This observation reveals that physical entanglement is directly required for logical magic. This can be understood by the fact that, while logical Pauli states such as $\ket{+_L}$ are eigenstates of operators $X_L =  X_1X_2X_3...$, which is a tensor product of physical operators, states such as $\ket{T_L}$ are eigenstates of $\frac{1}{\sqrt{2}} (X_1X_2X_3... + Y_1Y_2Y_3...)$, involving a superposition spanning the code that is necessarily entangled. Such an observation has fundamental implications: for example, an error-corrected Bell inequality test (involving T-basis measurements) requires a high degree of in-block entanglement. In ED Fig.~\ref{fig:ED_Universality}f we perform such an error-corrected Bell inequality test and measure a CHSH inequality of $1.99(3) \times \sqrt{2}$, saturating the quantum bound \cite{Bell1964}.

\subsection*{Deep circuits at constant entropy}

\begin{figure}
\includegraphics[width=1\columnwidth]{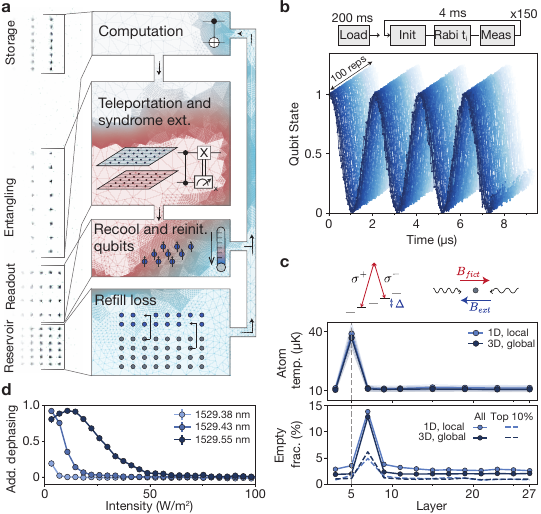}
\caption{\textbf{Architecture for constant-entropy computation.} \textbf{a,} Illustration of processes for removing entropy generated by computation. Logical teleportation is used to ensure all physical errors are removed (Fig.~6e). \textbf{b,} Rabi oscillations measured using the same atoms for 150 cycles of non-destructive measurement and re-initialization. Each curve shows a single experimental run, averaged over 200 atoms in parallel. 3D cooling methods are used in this subplot as coherence does not need to be preserved. \textbf{c,} Local cooling with 1D polarization gradient cooling (PGC) and electromagnetically induced transparency (EIT). The finite magnetic field is compensated by applying a relative detuning between the two circularly polarized counterpropagating beams, lending to a rotating frame where the effective field is zero. The atom loss and temperature is constant as a function of time and recovers to steady state within one cycle after applying a perturbation (turning off cooling on one layer) to the system (dashed line). \textbf{d,} Additional shielding of the data qubits is provided by a 1529-nm shielding laser, which rapidly suppresses decoherence induced by the imaging (resonance is at 1529.365 nm).}
\label{fig5}
\end{figure}

\begin{figure*}
\includegraphics[width=2\columnwidth]{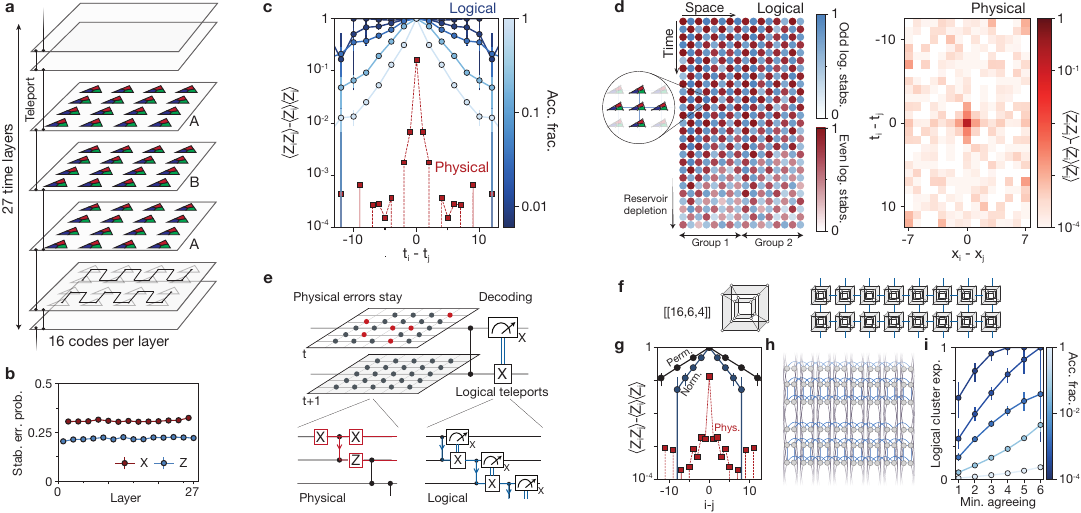}
\caption{\textbf{Deep logical circuits at constant entropy.} \textbf{a,} Schematic of a 2D cluster state created from Steane codes in space and time. The same atomic qubits are re-used every other time layer for up to 27 layers. \textbf{b,} Results of repeated logical state preparation of Steane codes. Stabilizer error is constant as a function of cycle. \textbf{c,d,} Logical and physical correlations in 1D (\textbf{c}) and 2D (\textbf{d}) cluster states. Logical correlations persist while stabilizer error correlations rapidly decay. Here and below, various degree of acceptance fraction are applied (see Methods for tabulation of all details). \textbf{e,} Circuit diagram illustrating that logical-level evolution is unitary and allows the logical operator to propagate throughout the algorithm whereas physical-level evolution is dissipative and does not let physical errors propagate. \textbf{f,} High-rate [[16,6,4]] codes are entangled in space and time direction. \textbf{g,} Logical evolution and physical error removal with [[16,6,4]] codes (1D cluster state in time direction). Permutation CNOTs within the block (black curve) extend the correlation length in comparison to non-permuted qubits (Methods). \textbf{h,} Entanglement structure in space and time, with up to 96 $d=4$ logical qubits active simultaneously. \textbf{i,} Logical 2D cluster expectation values (averaged across space and the first 9 teleportation layers) as a function of the number of co-propagating logical operators which agree (i.e., cluster state stabilizers have the same outcome for each logical qubit within the block).
}
\label{fig6}
\end{figure*}

We now explore the ability to perform deep-circuit quantum computation on the logical level. A critical requirement is that the processor is kept at a constant entropy (Fig.~5a) \cite{Aharonov1999}, necessitating that physical errors do not accumulate. This is challenging because computation inevitably introduces errors in the physical qubit state while, in addition, increasing entropy in the other degrees of freedom, such as the atomic motional state. To ensure that all physical errors are removed and that computation is kept at constant entropy, we again leverage transversal teleportation \cite{Raussendorf2001,Knill2005, Raussendorf2012,Sahay2023a}. In this approach (detailed below) the logical information propagates throughout the circuit while physical errors are left behind. Measuring this block then enables qubit reset, re-cooling, and re-initialization of the physical atoms. After that the block is prepared again in a low entropy state and is then utilized for subsequent teleportation steps. 

To realize constant entropy operation during deep quantum circuits, the atomic internal states, temperature, and atom filling need to be re-initialized during the computation. In order to achieve this, we combine our non-destructive internal state readout (Fig.~1b) with non-destructive imaging that also re-cools the atom. While laser cooling is typically achieved using 3D beams in zero magnetic field, we implement a novel method that enables high-fidelity imaging and cooling operating with focused 1D beams in a finite B-field (required for atomic qubit control) (Fig.~5c) \cite{Rolston1992,Chow2024}. We further protect coherence of data qubits in the nearby storage zone by applying a 1529-nm shielding beam (Fig.~5d, ED Fig.~\ref{fig:ED_1529})~\cite{Hu2024}. Moreover, the missing atoms in the array are refilled with atoms from the atomic reservoir (Fig.~5a). With all these methods combined, in Fig.~5c we measure the performance when subjecting the atoms to all the operations (except entangling gates) in a 27-layer circuit, explained below. For instance, by applying a perturbation on the 5th cycle, we find that the atomic filling and temperature quickly recover to a steady-state. We find that the 1D cooling methods nearly reproduce the conventional 3D performance, limited by tweezer-depth inhomogeneity (which can be straightforwardly improved, see ED Fig.~\ref{fig:ED_ImagingAndCooling}). Repeated operation also allows for fast cycle rates; as an example, Fig.~5b shows Rabi calibrations with a 4-ms cycle time.

We now use these tools to implement logical algorithms. Here, we entangle, measure, and teleport the logical blocks in alternating A and B groups (Fig.~6a). Within one layer, we bring in a fresh batch of physical qubits (group A) from the readout into the entangling zone in order to encode them into error-correcting codes, and then entangle the code blocks with each other (entangling in space direction).  We then bring group B (already entangled) from storage, and perform a transversal entangling gate with the group A block (entangling in time direction). In the spirit of measurement-based quantum computing \cite{Raussendorf2001}, we then move group B into the readout zone, group A to the storage zone, and then measure group B. Group B is then re-initialized and the whole process is repeated in the next layer. 

We first study repeated state preparation of 32 blocks of [[7,1,3]] (Steane) codes (16 blocks per alternating group) for up to 27 layers (Fig.~6a), and find that the stabilizer expectation value remains constant as a function of the cycle number (Fig.~6b), indicating a steady-state internal entropy. (We note that in these experiment the fidelity is limited by several sub-optimal choices in the circuit design structure, see Methods for details). As an example algorithm, we entangle the [[7,1,3]] codes in both the time and space directions to realize 1D and 2D cluster states and probe the resulting correlations \cite{Raussendorf2001}. Starting with a 1D cluster state in the time direction, Fig.~6c shows the correlations on both the logical-level and the physical-error-level by evaluating the correlator $\langle Z_i Z_j \rangle  - \langle Z_i \rangle \langle Z_j \rangle $ between coordinates separated in time (Methods). On the logical-qubit level we observe the expected correlations between logical outcomes corresponding to successful algorithm evolution, with a decay with distance corresponding to an effective algorithmic error rate that improves with decreasing entropy (corresponding to lower acceptance fraction). Conversely, for the physical errors, we observe the correlations are rapidly suppressed. Similar behavior is found in the 2D cluster state data (Fig.~6d), where the logical cluster state stabilizers are finite across both space and time, but the correlations between stabilizer errors are rapidly suppressed. These properties persist until the reservoir begins to run out of atoms. Leveraging the regular transversal measurements, we use data qubit loss detection as well as a recurrent neural network architecture \cite{BonillaAtaides2025} for decoding these algorithms (Methods).

These observations can be understood by considering the processes illustrated in Fig.~6e, indicating how the transversal teleportation ensures the removal of physical errors. Whereas the decoding and logical teleportation (with feedforward done in-software) maintains unitary evolution and propagates the logical information, the physical errors remain on the previous block (with errors propagated by entangling gates traveling at most one layer). This structure, where the logical information is natively teleported onto a fresh logical block in the algorithm, thereby ensures that the algorithm proceeds at constant entropy while also performing logic (also leveraged in Fig.~4). 

To test this teleportation method for more general encodings, we study high-rate [[16,6,4]] codes \cite{Reichardt2024a}. Such high-rate codes \cite{Bravyi2024} have a complex structure that would generally require intricate circuit structuring to ensure leakage removal \cite{Acharya2024, Baranes2025}. In contrast, by directly using the teleportation procedure described above we find (blue curve in Fig.~6g) that in a temporal 1D cluster state the logical information propagates while the physical errors are suppressed with a rate similar to 
the simpler [[7,1,3]] Steane codes. Such high-rate codes also enable new opportunities for realizing quantum algorithms. For example, the permutation CNOT operation can enable entangling logical qubits within the same block simply through re-indexing physical qubits, thereby extending the correlation length (black curve in Fig.~6g). 

We also explore 2D entanglement of the [[16,6,4]] blocks (with up to 16 blocks at a time), realizing the entangled structure depicted in Fig.~6h. Fig.~6i shows the 2D logical cluster state stabilizers as a function of postselection on shots where the co-propagating logical operators agree (i.e., the cluster state stabilizers have the same outcome for each logical qubit within the block). We find that such a procedure further improves algorithm performance. This is because the logical operators - although independent degrees of freedom - are supported on the same physical qubits. Interestingly, while these easily attained in-block correlations are algorithmically useful (as illustrated using [[8,3,2]] codes in Ref.~\cite{Bluvstein2023}), generating such logical entanglement is only possible because the logical operators overlap, which is enabled by the underlying physical entanglement (Methods).

\subsection*{Discussion and outlook}

We now turn to a discussion of our key observations. 
First, our experiments uncover an intricate  interplay between quantum logic operations and entropy removal: understanding their relationship enables syndrome extraction to be applied only where necessary (Fig.~1-6), and also clarifies entropy-related aspects of logical gate performance (Fig.~3). Second, we find that physical entanglement should be deployed judiciously in fault-tolerant systems.  While techniques such as transversal gates and high-rate encoding reduce the physical entanglement required for a given amount of logical operations (Figs.~3,6f-i), logical magic states demand enforcing precise entanglement structure (Fig.~4). Third, we observe that logical teleportation can be a central mechanism for FTQC \cite{Knill2005}.  Such a teleportation enables universality even with fully transversal operations (Fig.~4) and offers a native path to physical error removal in deep-circuit protocols, including those involving high-rate codes, allowing  direct data qubit detection (including atom loss events) to be effectively used in the decoding algorithms (Figs.~2,6). Beyond these architectural insights, we also observe several fundamental physics aspects of QEC. While prior work has connected quantum contextuality (involving $T$-basis measurements) with computational hardness \cite{Howard2014}, our results suggest  additional links to the essential entanglement required for the QEC (Fig.~4 and ED Fig.~\ref{fig:ED_Universality}). These observations point toward a deeper understanding of how to protect algorithmic outputs, and facilitate the realization of complex, deep-circuit quantum algorithms.

While the current experiments demonstrate QEC performance a factor of $\approx 2$ below key thresholds, large-scale computation will greatly benefit from reducing physical errors (tabulated in Fig.~2f). Based on our observations, we estimate that an additional 3-5 fold physical error reduction can be achieved by direct  improvements in single-qubit operations, improved  calibration (Fig.~5b), and a $4$x increase in entangling laser power \cite{Evered2023a} (Methods). While machine learning decoding is found to be both effective for entropy removal performance (Fig.~2) and simple algorithms (Fig.~6), and fast ($\sim1 ~\mu$s per shot, batched on a GPU (Methods)), more work is required to ensure scalability of this powerful method \cite{Bausch2024}. Moreover, although the maximum circuit depth here is limited by a depleting atomic reservoir, in a complementary experiment conducted in a separate apparatus, coherent continuous operation on over 3000 atomic qubits \cite{Chiu2025} is demonstrated via continuous atom reloading \cite{Gyger2024,Norcia2024}, with techniques fully compatible with the methods presented here. Taken together, these techniques enable advanced experimental exploration of fault-tolerant universal algorithms. Combined with other significant progress with neutral atom systems \cite{Jenkins2022,Ma2023a,Manetsch2024,Tsai2025, Muniz2025a, Zhang2025a}, such developments demonstrate that these systems are uniquely positioned for experimental realizations of deep-circuit fault-tolerant quantum computing.

\clearpage
\newpage

\bibliographystyle{naturemag_arxiv2.bst}
\bibliography{library.bib}

\clearpage
\newpage

\section*{Methods}

\small \noindent\textbf{System overview} \\

To carry out the present experiments, several key upgrades have been made. We provide here an overview of the experimental system (ED Figure~\ref{fig:ED_systemdiagram}).

A cloud containing millions of cold $^{87}$Rb atoms is loaded in a magneto-optical trap inside of a glass vacuum cell. The Rb atoms are then loaded stochastically into programmable, static arrangements of 852-nm traps generated with a spatial light modulator (SLM, Hamamatsu X13138-02), and then rearranged with a set of 852-nm moving traps generated by a pair of crossed acousto-optic deflectors (AODs, DTSX-400, AA Opto-Electronic) to realize defect-free arrays~\cite{Barredo2016, Scholl2021, Ebadi2021}. We use D1 lambda-enhanced gray molasses cooling to achieve a loading efficiency of 75\% \cite{Brown2019b}. Atoms are imaged with a 0.65-NA objective (Special Optics) onto a CMOS camera (Hamamatsu ORCA-Quest C15550-20UP), chosen for fast electronic readout times. The qubit state is encoded in $m_F = 0$ hyperfine clock states in the $^{87}$Rb ground-state manifold, with $T_2 >1$s~\cite{Bluvstein2022, Graham2022}, and fast, high-fidelity single-qubit control is executed by two-photon Raman excitation~\cite{Bluvstein2022,Levine2021}. A global Raman path illuminating the entire array is used for global rotations (Rabi frequency $\sim $0.5 MHz, resulting in $\sim 5 ~\mu$s rotations with composite pulse techniques~\cite{Bluvstein2022}) as well as for dynamical decoupling throughout the entire circuit (typically 1 global $\pi$ pulse per movement). For this work, we upgrade our microwave source (Rohde and Schwarz, SMW200A) and increase our intermediate-state detuning to 550 GHz (measured scattering error $5\times10^{-5}$ per robust SCROFULOUS pulse).
Fully programmable local single-qubit rotations are realized with the same Raman light but redirected through a local path which is focused onto targeted atoms by an additional set of 2D AODs. To realize high-fidelity, programmable single-qubit pulses we have made upgrades to our single-qubit addressing to use direct Raman X-type rotations (see Raman gates section). Entangling gates (270-ns duration) between clock qubits are performed with fast two-photon excitation using 420-nm and 1013-nm Rydberg beams to n=53 Rydberg states, utilizing a time-optimal two-qubit gate pulse~\cite{Jandura2022} detailed in Ref.~\cite{Evered2023a}, in this work with the 420-nm laser red-detuned by 4.8 GHz from the intermediate state. During the computation, atoms are rearranged with the AOD traps to enable arbitrary connectivity~\cite{Bluvstein2022}.
An important upgrade in this work is the ability to perform non-destructive qubit readout, enabling loss detection as well as qubit re-use. We realize this with a one-dimensional optical lattice \cite{Wu2019} which pins one of two spin states, use optical tweezers to separate the pinned and unpinned states, and then image the atom position. To further enable mid-circuit qubit measurement and re-use on large arrays, we develop methods of low-loss, high-fidelity qubit readout and re-initialization while only needing moderate trap depths (see below). We use these techniques here for re-using atoms and extending the depth of error-corrected computation. \\

The quantum circuits are programmed with a control infrastructure consisting of five arbitrary waveform generators (AWG) (Spectrum Instrumentation), as illustrated in ED Fig.~\ref{fig:ED_systemdiagram}b, synchronized to $<$ 10-ns jitter. The 2-channel rearrangement AWG is used for real-time rearrangement, the 2 channels of the Rydberg AWG are used for entangling gate pulses and for local SLM detunings, the 4 channels of the Raman AWG are used for IQ (in-phase and quadrature) control of a 6.8 GHz source~\cite{Levine2021,Bluvstein2022} (the global phase reference for all qubits) and pulse-shaping of the global and local Raman driving, the 2 channels of the Raman AOD AWG are used for displaying tones that create the programmable light grids for local single-qubit control, and the 2 channels of the Moving AOD AWG are used for controlling the positions of all atoms during the circuit. 

In this work, we realize circuits as long as 1.1 seconds for the experiments in Figures 5 and 6. In order to realize this with the AWGs, we generate a memory segment for one circuit layer for the Moving AWG, Rydberg AWG, and Raman AOD AWG, and then loop these identical memory segments for each layer. This is complicated for the Raman AWG as phase continuity needs to be ensured, and so for simplicity in this work we program the whole Raman waveform directly. We fill the entire memory of the Spectrum AWG and this is what limits our experiments to 27 layers here (and then we choose an appropriately sized reservoir to have atoms for that many layers). Future work will benefit significantly from improved waveform streaming. \\

\small \noindent\textbf{Details of processor configuration} \\

Our approach to quantum processing is highly programmable, however we find that each new atomic layout design behaves slightly differently \cite{Bluvstein2022,Bluvstein2023,Evered2025}. A close analogy here is that we can ``design and print a new chip'' every time we change our processor design, but each one requires its own specific characterization and calibration. We observe that - while each configuration we create can be slightly different and can have its own specific challenges - with sufficient characterization and optimization we can recover `nominal' performance (i.e., consistent with simple single-qubit and two-qubit error model), and that such a configuration is stable and reproducible once it has been properly set up.

For example, we detail some example circuit configurations that required different degrees of characterization in this work. In the repeated QEC rounds on the surface code, we were careful to engineer the circuit structuring in a manner where the time would perfectly echo on each qubit. This was greatly facilitated by the symmetric four-gate structure of the stabilizer syndrome extraction circuit. For example, although the local Raman pulses are applied row-by-row, we ensure that the overall amount of time in superposition - although different for each atom - echoes around a central global $\pi$ pulse. However, although this enabled us to ensure the total time echoed, the specific structuring and parity of pulses prevented us from ensuring the overall atomic trajectory echoed \cite{Bluvstein2023}. As such, we had to be more careful with homogenizing the AOD trap power over the surface code region. Conversely, in order to realize the programmable hypercube codes, we confined ourselves to the general encoding circuit where even the total time did not echo on each qubit, which thereby greatly impacted performance. These illustrate that each circuit we realize is different and, although we find it is always possible to achieve correct, `nominal' fidelities, sometimes our layout and circuit design requires multiple iterations to find a suitable approach. 

We now detail some more specific aspects of the processor designs used in this work. \\

\noindent{\it Surface code.}
For surface code experiments (Figs.~1-3), the same static traps are used for mid-circuit storage of ancilla blocks and for readout of all qubits at the end of the computation. The readout zone is 12 rows tall ($55\,\mu m$) with two rows of traps per atom for the lattice readout, Fig.~2a. Six blocks of qubits are interlaced horizontally for storage, corresponding to five (four) 6x6 ancilla blocks and one (two) 5x5 data block(s) in Fig.~2(3). This interlacing ensures that the dimensions of each qubit block is the same in both the readout and entangling zones, preventing heating from AOD intermodulation effects that we observe when compressing or expanding the AOD grid.
Two additional columns of traps form a small reservoir used for initial rearrangement, resulting in a total array width of $165\,\mu m$. 

The 420-nm and 1013-nm Rydberg tophat beams cover 7 rows of gate sites in the entangling zone and are homogenized to $\approx$1\% peak-to-peak variation over a vertical extent of $60\,\mu m$. The entangling zone is separated by $40\,\mu m$ from the storage and readout zone (overlapping in these measurements) to ensure negligible error on stored qubits from the tails of the Rydberg beams. 
\\

\noindent{\it Deep circuits.} 
For deep circuit experiments (Figs.~5-6, and same configuration used in Fig.~4), we choose the same $60\,\mu m$ vertical extent for the entangling zone as above. Within this zone, entangling gates are performed simultaneously on up to 256 qubits across 8 rows and 16 columns of gate sites with a horizontal extent of $175\,\mu m$. Below the entangling zone is the readout zone, used for measurement and re-initialization of up to 128 atoms arranged in 4 rows. This region is illuminated by counterpropagating imaging and cooling beams (beam waist $50\,\mu m$) as well the one-dimensional lattice beams (average waist $60\,\mu m$) as illustrated in ED Fig. \ref{fig:ED_systemdiagram}d. 

During mid-circuit imaging, atoms are always held in the storage zone, 50 $\mu$m from the entangling zone. To preserve coherence of qubits during the imaging, the storage zone is illuminated by a 1529-nm shielding beam with a beam waist of $35\,\mu m$, matching the zone's vertical extent. These design parameters ensure both that stored atoms do not pick up error from Rydberg beams, as described above, and also that negligible 1529-nm light reaches the readout zone and so does not cause spurious lightshifts on the imaging and cooling transitions (see ``1529-nm shielding beam''). Finally, the reservoir is located directly below the readout zone and contains up to 196 atoms in 6 rows.

The trap intensities in the entangling and storage zones are set to half of those in the readout and reservoir zones to improve qubit coherence. This is achieved by modifying the target trap intensities in the trap generation algorithm \cite{Ebadi2021}. We center the array on the zeroth diffraction order of the trap SLM to maximize the deflection efficiency.

In our first attempt of deep-circuit processing we made multiple processor design decisions that are suboptimal and impacted our fidelity, which we list here. There is no fundamental reason for these, and after the first circuits implemented here these can be readily improved for future experiments.

\begin{itemize}
    \item Re-initialization with local Raman led to an overly sensitive re-initialization procedure that complicated calibration.
    \item Imperfect echoing due to lack of symmetry in the generalized hypercube encoding circuit led to high sensitivity to trap depth variations.
    \item We shifted the trap path and this led to an exacerbated AOD intermodulation effect that appears to cause significant heating and a reduced $T_1$.
    \item Our SLM array had trap depth inhomogeneity, affecting cooling performance (Fig.~5c) and exacerbating improper echoing issues.
    \item The specific Rydberg tophat beams we used here had an exacerbated inhomogeneity of $\approx 2-3\%$ peak-to-peak variation.
    \item We found that performance is sensitive to the 1529-nm beam profile, requiring homogeneous coverage in the storage zone due to complex resonances, while preventing illumination on the readout zone.
    \item Magnetic field noise from the current supply impacted coherence preservation during between-layer idle times. 
\end{itemize}

\small \noindent\textbf{Spin-to-position conversion with a one-dimensional optical lattice} \\

We realize non-destructive qubit readout throughout this work through spin-to-position conversion (ED Figure \ref{fig:ED_Lattice}) \cite{Mandel2003, Weiss2004, Robens2017a, Wu2019}.
A one-dimensional optical lattice is formed by two 795-nm counterpropagating local beams, both sourced from the same titanium:sapphire laser (M Squared) and operated at 50-200~GHz blue-detuned of the D1 line. Both beams are $\sigma^-$ polarized such that $\ket{F=2;m_F = -2}$ is a dark state and $\ket{F=2;m_F = +2}$ experiences a maximum lightshift of approximately 6 MHz, corresponding to approximately 300 kHz trap frequency in one axis. The close detuning is a balance between minimizing off-resonant coupling to the D2 line for the dark state and reducing scattering and heating from the lattice light.
Since the clock state qubit is used for computation, for readout we first optically pump $\ket{F=2;m_F=0}$ into the dark state with 780-nm $\sigma^-$-polarized light resonant to $F=2$ to $F'=3$ which is co-propagating with one port of the lattice. To suppress the probability of scattering into the dark state during readout, we further transfer $\ket{F=1;m_F=0}$ to  $\ket{F=2;m_F=+2}$ (bright state), which also increases the trap depth. This is achieved via either a coherent Raman transfer or with incoherent $\sigma^+$-polarized 780-nm repumper from $F=1$ to $F'=2$. We use the former approach in all surface code experiments (Figs.~1c, 2-3) and the latter in deep circuit experiments (Figs.~1b, 4-6), finding comparable performance from both methods.

Following these state transfers, the lattice is ramped up adiabatically over approximately 100 $\mu$s. AOD tweezers pick up atoms in the dark state and move them by approximately $2\,\mu m$ over approximately 500 $\mu$s; during this, atoms in the bright state are pinned in place by the stronger confinement of the lattice. Finally, the lattice is ramped down and conventional camera-based readout then images the position of the atom, allowing identification of the spin state as well as loss detection. Using the data in Fig.~1b, we measure an error probability of 0.87(7)\% for the dark state, 0.05(5)\% for the bright state, and a 0.24(2)\% probability of loss. The asymmetric error arises from trade-offs when simultaneously optimizing for loss and readout fidelity and can be tuned to be more balanced; typically, due to the pumping fidelity, the dark state error is at least $\sim 0.3\%$ higher than the bright state.\\

\small \noindent\textbf{One-dimensional and finite-field operation for imaging and cooling} \\

For local cooling and imaging, we use two counterpropagating 780-nm beams with opposite circular polarization \cite{Dalibard1989,Rolston1992} (ED Figure~\ref{fig:ED_ImagingAndCooling}). The beams are red-detuned from $F=2$ to $F'=3$ and have a variable relative detuning; the $\sigma^+$-polarized beam additionally contains a small repump component. Conventional methods based on polarization-gradient cooling (PGC) require zero magnetic field, however mid-circuit operation requires a finite magnetic field to maintain the quantum state of active qubits. To this end, we develop a scheme for one-dimensional PGC in finite magnetic field. PGC is based on a linear polarization rotating along the beam propagation direction which produces a population imbalance within the hyperfine levels \cite{Rolston1992}; in finite field, this imbalance is disturbed and the cooling mechanism breaks down \cite{Walhout1992}. By transforming to a frame where the polarization rotates in time, a fictitious field appears which cancels the external field and restores the cooling effect. This condition is achieved by detuning the two counterpropagating beams - which are (anti)parallel to the external magnetic field - by two times the Zeeman splitting of adjacent $m_F$ levels. As shown in ED Fig.~\ref{fig:ED_ImagingAndCooling}e, this detuning method works across the full range of magnetic fields studied (up to 8.6~G).
Furthermore, it is broadly applicable to finite-field implementation of any one-dimensional technique based on the same polarization configuration used here, for example gray-molasses cooling \cite{Grier2013, Brown2019b}.

While this finite-field PGC is sufficient to image without loss, we add a second stage of EIT cooling to further reduce the atom temperature \cite{Chow2024}. The scheme, shown in ED Fig.~\ref{fig:ED_ImagingAndCooling}f, uses the same beams as the PGC imaging and only requires changing to be blue-detuned of $F=2$ to $F'=2$ (by $\sim$80~MHz) and reducing the power in one of the beams. Since the cooling is uniaxial, it is \textit{a priori} unclear if all three motional degrees of freedom can be cooled with such techniques. Using both drop-recapture measurements and adiabatic ramp-down measurements of the atom temperature \cite{Tuchendler2008}, we probe the radial and axial atom temperature and find them both to be comparable to three-dimensional techniques (Fig.~5c and ED Fig.~\ref{fig:ED_ImagingAndCooling}f).
Furthermore, the steady-state temperature and loss is set only by the EIT cooling fidelity and is independent of the degree of heating introduced from the prior circuit. \\

\small \noindent\textbf{1529-nm shielding beam} \\

To preserve the coherence of qubits in the storage zone, we illuminate them with a single beam of 1529-nm light (ED Figure~\ref{fig:ED_1529}). By coupling the $5P_{3/2}$ state to the $4D_{5/2}$ state, we impart a strong Stark-shift on the excited $5P_{3/2}$ state \cite{Hu2024}. This causes probe light in the readout zone to appear off-resonant to the storage zone atoms while maintaining qubit information in the hyperfine manifold of the ground state, see ED Fig.~\ref{fig:ED_1529}a.
The beam is generated by a Connet CoSF-D series 10W fiber laser and is focused down to an elliptical waist of $35~\mu m \times 65 ~\mu m$. The shorter waist of the beam is aligned vertically to the center of the storage zone. We image the beam in a 4f system and apply a knife-edge in the image plane, approximately 4 beam waists from its center, to suppress its Gaussian-tail. Stray 1529-nm light, even at low powers, can degrade the imaging quality in the readout zone. We find, therefore, that beam-shaping is critical for maintaining stable imaging quality and coherence on the storage-zone atoms for the layout of our array.

 We measure dephasing of the storage-zone qubits, as a function of detuning from the bare transition, while readout-zone qubits are illuminated with local probe and repumper light, ED Fig.~\ref{fig:ED_1529}b. We capture the key features of the spectrum with a simple model where the additional dephasing at each drive power scales as 
      $\exp{ \left( -\frac{\Omega^2_\textrm{probe}\Gamma_\textrm{probe} t}{4 \Delta_{LS}^2} \right)}$
 where $\Omega_\textrm{probe}, \Gamma_\textrm{probe}$ are the Rabi frequency and scattering rate of the local imaging beams, $t$ is the illumination time, and $\Delta_{LS}$ is the calculated lightshift of $5D_{3/2}$ due to the coupling to $4D_{5/2}$ and $4D_{3/2}$. More complex on-resonance or multi-level features are not captured by this simple model and are particularly sensitive at detunings between the resonances of the $4D$-levels \cite{Sibalic2017}. During all experiments with qubit re-use and local imaging, we address the storage zone at 1529.49\,nm with $\approx 1.2$\,W. This corresponds to an approximate lightshift of 6\,GHz on the $5P_{3/2}$ state. To further characterize the 1529-nm laser we explore varying the detuning of the local imaging light and observe a clear Autler-Townes splitting, ED Fig.~\ref{fig:ED_1529}c. We find that, as expected, the separation of two fitted Lorentzian peaks scales linearly with the square-root of the drive power.\\

\small \noindent\textbf{Repeated rearrangement from reservoir} \\

The mid-circuit image identifies the qubit state as well as which atoms are lost. Before rearrangement, the atoms are recombined into their original tweezers, balancing the trap depth between the AOD and SLM tweezers to minimize loss and using cooling throughout. After this recombination, we fill empty sites using the reservoir.

In each round of rearrangement, target rows are refilled sequentially with one parallel step per row. All atoms in each step are sourced from a single reservoir row. We choose efficient horizontal moves and optimize the reservoir-to-target row pairings to minimize travel distance. Finally, since the local imaging beams do not cover the full extent of the reservoir, the reservoir site occupancies are stored from a global image before the circuit begins and used reservoir atoms are tracked in software. This leads to a slowly growing rearrangement infidelity.\\

\small \noindent\textbf{Mid-circuit re-initialization} \\

After qubits have been measured and atom loss refilled, the spin state is re-initialized to re-use the qubit. This local state preparation is performed in the readout zone using a Raman-assisted optical pumping scheme \cite{Bluvstein2022, Levine2019}. Local Raman is used for the coherent $\pi$-pulses and the local probe beams are used for resonant depumping of the $F=2$ manifold. Due to the close horizontal spacing of traps in the readout zone, we minimize crosstalk between local Raman tweezers by alternating the applied $\pi$-pulses between odd and even columns. We perform 24 cycles of pumping per atom over a few hundred microseconds.\\

\small \noindent\textbf{Local single-qubit gate details} \\

Single-qubit gates are performed using Raman transitions as previously described in Ref. \cite{Bluvstein2023}, with several changes to allow $X(\theta)$ rotations to be directly implemented with high fidelity. The key challenge for local $X$ gates is ensuring polarization homogeneity, since the Rabi frequency is sensitive to the degree of circularity. We find inhomogeneity both across the array, introduced by a sharp dichroic cut-off noted in ref. \cite{Bluvstein2023}, as well as inhomogeneity within each optical tweezer due to polarization breakdown near the tweezer focus.
To reduce the first effect, we add a second copy of the dichroic into the path with a half-waveplate between the pair, such that any angle-dependent phase shifts upon reflection from the dichroics are equally applied to both the s- and p-polarized components and the polarization remains close to circular.
Second, polarization breakdown of a circularly-polarized tweezer results in an off-axis fictitious field with components both parallel and perpendicular to the external magnetic field (in the plane of the tweezer focus) \cite{Mukunda1987}; the parallel components can drive Raman transitions and result in dephasing of the clock qubit. Since the magnitude of the maximum off-axis field falls off linearly with tweezer waist, we mitigate this by increasing the waist to $2.5\,\mu m$. Finally, to increase the projection of the Rabi frequency drive along the magnetic field axis, we displace the Raman beam by roughly $1.5$ mm within the back aperture of the objective whose size is 5.5 mm, so that the Raman beam comes in at an angle. For all single-qubit gates in this work, we use robust SCROFULOUS pulses \cite{Cummins2003}. \\

\small \noindent\textbf{AOD intermodulation effects} \\

We observe several intermodulation effects from the AODs that can result in degraded performance for specific AOD moves. First, it is important to ensure that the frequency tones in a given AOD axis are in an exact frequency comb, as intermodulation can lead to interference and beating near trap frequencies. Second, we observe here that the relative frequencies of the X-frequency-spacing and Y-frequency-spacing is also important, and that when beat notes of these are near trap frequencies this can also lead to heating. As such, we now primarily use the AODs for translations, avoiding compressions / expansions of the grid when possible, and choose incommensurate spacings for X and Y to avoid accidental cross-resonances. \\

\small \noindent\textbf{Analysis of error correlations} \\

Correlations in errors, in either space or time, can have important implications on quantum error correction. Here we explain various correlation analyses in our system. \\

\noindent{\it De-correlation of global coherent errors by projective measurement.} Parallel control enables us to, for example, realize a transversal entangling gate with a single global pulse of our entangling laser~\cite{Bluvstein2023}. One may be concerned that such a global control can lead to globally correlated errors that can affect error correction performance. However, error correction natively de-correlates such errors.

Consider a code block of qubits with X and Z stabilizers. Applying a global $\theta$ will map each of the X operators to $\rightarrow (X+i\theta Y) = X \cdot (1- \theta Z)$. Consequently, measuring the X-basis component of this qubit will probabilistically lead to a Pauli $Z$ error on this site with probability $\theta^2$. Note that, for global rotation $\theta$, the logical operator $X_L = XXXX...$ maps to $ \rightarrow (X+i \theta Y)(X+i \theta Y)(X+i \theta Y)(X+i \theta Y)... = XXXX... + i\theta YXXX ... + (i \theta)^{d}YYYY...$. As such, for small $\theta$, logical rotations are exponentially suppressed with the code distance $d$. As such, even though all the physical qubits receive a global rotation $\theta$, the logical qubit state does not receive that same rotation, and after syndrome measurements these errors are converted into incoherent-type errors and can be corrected. This is the basis behind the observed suppression in Figure~1. In ED Figure~\ref{fig:ED_Coherent}b, we further show that the error correction prevents an unintended logical rotation. The logical rotation here is even further suppressed by the random stabilizer signs (below).\\

\noindent{\it Role of stabilizer signs under coherent errors.} 
Stabilizer signs affect the logical qubit's response to global coherent rotations. For transversal non-Clifford gates, deterministic stabilizer eigenvalues (e.g. = +1) are necessary to correctly implement the logical gate (Fig.~4). In contrast, Clifford circuits allow the eigenvalues to be either +1 or $-1$, as the signs can be simply tracked through the circuit, giving freedom to engineer how coherent errors interfere. For example, choosing negative signs can generate decoherence-free subspaces \cite{Debroy2021} and give greater robustness of the logical operator against coherent errors.  
The same principle also suppresses logical coherent errors during computation \cite{Bravyi2018New}. In particular, stabilizer measurement projects the logical state onto a specific stabilizer configuration with random $\pm$ 1 values, which corresponds to a random configuration of physical $X$ and $Z$ flips which do not commute with the coherent rotation. On top of the exponential suppression of logical coherent errors, this further results such that the specific rotation angle of scale $\sim \theta^d$ is random on each shot, effectively turning these again into incoherent errors on the logical level.\\

\noindent{\it Decay to Rydberg P states.} In Ref.~\cite{Evered2023a} ED Figure~7, we analyzed the presence of weak correlations between CZ gate errors seen in a repeated randomized benchmarking sequence, and speculated the origin of these may be due to decay to atomic Rydberg P states. Concretely, during Rydberg gates, roughly $0.07\%$ of the error budget is decay of Rydberg atoms to adjacent Rydberg P states. These states have a strong, long-ranged interaction with the Rydberg $S$ states that are used for the gates, and can thereby affect gates occurring in a different site, and moreover can have lifetimes of over 100 $\mu$s. During repeated benchmarking sequences, like the ones reported in Ref.~\cite{Evered2023a}, we have only $4~\mu$s between gates, and consequently Rydberg P atoms can survive for many layers of gates and corrupt gates in distant sites. 

In ED Figure \ref{fig:ED_ExtraSurface}g, we plot the CZ gate fidelity in a repeated benchmarking sequence as a function of the duration between the gates, and find that the gate fidelity in this array increases from 99.3\% to 99.5\% by increasing duration between gates to $100~\mu$s, as the Rydberg atoms decay or eject during that time. This also implies the reported gate fidelity in Ref.~\cite{Evered2023a} may have been impacted by this effect and thereby underestimating the maximum gate fidelity. Analogously, we observe that this gate fidelity reduction is removed by reducing atom density. In quantum circuits based on atom motion, the duration between gates is sufficiently long (e.g., 400 $\mu$s for surface code repeated stabilizer measurements) for these Rydberg P states to decay or eject \cite{Cao2024a}, which natively fixes this issue, and consequently we do not observe such effects during quantum circuits. \\

\noindent{\it Surface code measurements.} In our repeated quantum error correction on the surface code, we search for unexpected error correlations by plotting the distribution of detector errors in a shot. Crucially, we find these are closely consistent with the expected distribution as seen by Clifford simulations that assume uncorrelated one- and two-qubit errors. Although this data is only composed of approximately $10^5$ detector rounds (14855 shots, 96 detectors per shot across 5 rounds), it is nevertheless indicative of the absence of such events. We have yet to observe error burst events such as those observed in solid-state systems \cite{Acharya2024}. \\

\noindent{\it Logical teleportations and deep-circuit measurements}.
In a physical system, diverse physical errors and imperfections can cause complex correlations. For instance, a leakage event can lead to complex correlations that - without its knowledge - can greatly affect QEC performance. As discussed in the next section, incorporating logical teleportations in an architecture can ensure such errors are removed. In Figure~6 of the main text we verify that such teleportations indeed rapidly remove errors and ensure that errors are not correlated in either time or space. It is important that the atomic qubits are re-initialized properly for this to work. In ED Figure \ref{fig:ED_deepcircuits}c,d we plot the correlations when turning off cooling and sorting and find that correlations in such a case do not rapidly decay. \\

\small \noindent\textbf{Loss detection for improved QEC} \\

\noindent{\it Leakage types with neutral atoms}. Leakage errors, which take the qubit out of the two-level computational subspace, are important to account for in error correction. The three dominant leakage errors with neutral Rubidium (or other alkali) atoms are:

\begin{itemize}
    
\item \textit{Loss events}. Loss events are when the atom is physically lost from the optical trap. Due to the blockade nature of the gate, doing a gate with a lost atom simply turns off the gate while still applying gate error (it is identical to the atom being in state $\ket{0}$ which is also dark to the Rydberg laser).

\item \textit{Leakage to other hyperfine states in the ground-state manifold.} In the limit of a large magnetic field, these states are off-resonant and behave the same as a lost atom (turning off CZ gates). However, they are not detected through loss detection. Moreover, in the practical operating conditions of 8.6 G, the level spacings of 6 MHz (compared to Rabi frequency of 4.6 MHz) mean that adjacent hyperfine states can still off-resonantly couple to the Rydberg state and could lead to repeated errors.

\item \textit{Leakage to Rydberg states.} Population left in the Rydberg manifold can affect subsequent gates, and can lead to large error correlations. For example, many-body Rydberg evolution in dense systems observes so-called avalanche errors where a macroscopic fraction of the system has an error \cite{Festa2022}. Crucially, in our approach with a low atomic density and several hundred microseconds between gates, the Rydberg atoms (theoretically) either decay to the ground state or are expelled from the tweezer. In this way, such Rydberg leakage either convert into an error within the computational subspace, a leakage into adjacent hyperfine states, or a loss event.

\end{itemize}

We observe that with several hundred microseconds between gates, effects of Rydberg leakage are not apparent, and during our repeated QEC data we observe that our leakage is at least 80\% loss (ED Fig. \ref{fig:ED_SurfaceLoss}b).\\

\noindent{\it Effect of loss during repeated QEC.}
Although losses simply turn off subsequent gates, these lead to distinct signatures that are important to account for in the QEC design. Whereas ancilla loss will be detected in the projective measurement, losing a data qubit corresponds to unknown loss of a degree of freedom from the system \cite{Stace2009,Barrett2010}. Without adjusting the stabilizer measurement pattern to account for such a loss, the ancilla qubits now are measuring operators which anti-commute with each other, and thereby leads to a `flickering' pattern around the lost data atom. This flickering pattern is akin to the expected behavior in a subsystem code \cite{Kribs2005}, and means that the flickering can continue for arbitrarily long times. Without accounting for the loss, this then appears as strong time correlations which we observe in ED Fig. \ref{fig:ED_SurfaceLoss}b. 
\\

\noindent{\it Erasure information and superchecks.} It is useful to detect atom loss for two reasons. First, knowing about the lost atom greatly enhances the decoding performance. While bit-flip and phase-flip errors can be inferred by stabilizers, direct detection of qubit errors - or so-called erasures - means that one already has direct information about where the errors are. Such erasure information can thereby greatly improve decoding performance \cite{Wu2022,Scholl2023a,Sahay2023a,Chang2024,Putterman2025, Baranes2025}. For example, while only (d-1)/2 Pauli-type errors can be corrected, up to (d-1) erasure-type errors can be corrected. We do not detect erasures as soon as they occur, instead we detect them at the final qubit measurement, constituting delayed-erasure information. 

Second, although lost atoms lead to anti-commuting stabilizer measurements and a flickering error pattern, these can be accounted for with the use of so-called superchecks, illustrated in ED Fig.~\ref{fig:ED_SurfaceLoss}a \cite{Stace2009,Barrett2010}. While individual stabilizer checks around a lost atom are anti-commuting, taking products of multiple checks creates superchecks which again commute with each other. We find in Fig.~2b of the main text that such superchecks are able to remove the sharp rise in detected error that occurs with increasing data loss. \\

\small \noindent\textbf{Decoding} \\

\noindent{\it MLE and error-model tuning.} 
To decode the surface code experiments in Figs.~1-3, we use the delayed-erasure MLE decoder described in Ref.~\cite{Baranes2025}, augmenting the MLE decoder developed in Ref.~\cite{Cain2024a} to leverage loss information. In particular, the MLE decoder takes as input the stabilizer measurements and the probabilities of the physical error sources in the circuit, and outputs the most likely combination of errors consistent with the syndrome. We construct the circuit error model using Stim~\cite{Gidney2021a} to initially contain information about the Pauli error sources in the circuit, then update it for each shot to reflect the detected atom losses. In particular, after an atom is lost, all subsequent gates are canceled, generating different potential errors depending on when the loss occurred. We therefore consider all potential locations a qubit loss could have originated (e.g., initialization, gates, movement, or idling prior to measurement), then add each of the resulting error patterns and their probabilities to the error model for that shot. Errors producing the same syndrome are combined into a composite error mechanism and their probabilities are correspondingly reweighted, as in Ref.~\cite{Gidney2021a}. Note that this process explicitly accounts for both propagated Pauli errors from the gate cancellations and the invalidation of stabilizers, which are handled using superchecks. 

To optimize the performance of the MLE decoder, we fine-tune the probabilities of different error sources in the circuit error model. In particular, we associate each physical operation with both a Pauli and loss error rate. The error probabilities in these channels are then treated as variables which we optimize using the covariance matrix adaptation evolution strategy~\cite{Hansen2019} in order to minimize the logical error rate on a dataset of approximately 10000 shots (different shots from the final dataset used for evaluating the fidelity).

To quantify the benefit of using loss information, in Fig.~\ref{fig2} the `bare MLE' decoder does not update the circuit error model based on the loss information, and assigns each loss event to a $\ket{0}$ measurement. We find the loss information improves the measured $d=3/d=5$ error ratio from $1.24(5)$ to $1.69(8)$. 

Finally, we quantify the confidence of the MLE correction for each shot by comparing the probability of the most likely error $p_0$ and the probability of the most likely error which gives the correction to the logical Pauli observable $p_1$ ~\cite{Hutter2014, Bombin2024, Smith2024, Gidney2025}. The more similar these two error probabilities are, the less confident the decoder is in its correction. We can therefore postselect on increasing $p_0/(p_0+p_1)$ to improve the accuracy of the results, which we use in Fig.~3d when studying repeated logical gates. 
\\

\noindent{\it ML decoder - surface code.} We employ a fully connected neural network to decode measurement outcomes from the Fig.~2 surface code experiment using machine learning (ML) \cite{Krastanov2017,Baireuther2018,Gicev2023,Bausch2024}. The decoding task is formulated as a supervised binary classification problem: the input features are measurement outcomes from the experiment, and the output is a label indicating whether the initial state was $\ket{0_L}$ or $\ket{1_L}$. The ML architecture is a fully connected feedforward network comprising four linear layers, each followed by batch normalization and a Gaussian Error Linear Unit (GELU) activation, as illustrated below:
\begin{lstlisting}[language=Python, numbers=none]
decoder = nn.Sequential(
    nn.Linear(input_size, 1024),
    nn.BatchNorm1d(1024),
    nn.GELU(),
    nn.Linear(1024, 512),
    nn.BatchNorm1d(512),
    nn.GELU(),
    nn.Linear(512, 256),
    nn.BatchNorm1d(256),
    nn.GELU(),
    nn.Linear(256, 1),
    nn.Sigmoid()
)
\end{lstlisting}
\noindent Training proceeds in three stages: raw training, ensembling, and fine-tuning.

\textit{Raw training ---} We begin by training the decoder on simulated data generated through circuit-level simulations that incorporate both Pauli and loss errors.
Measurement outcomes take values of $0, 1,$ or $2$, corresponding to the qubit being in the $\ket{0}$ state, the $\ket{1}$ state, or being lost, respectively. These are one-hot encoded, so the feature vector of a given shot is $3 \times \text{(\# of measurements)}$.
To create balanced training data, random software flips are applied with probability $1/2$ along the relevant logical operator, yielding ensembles of ${\ket{0_L}, \ket{1_L}}$ for the $Z$ memory and ${\ket{+_L}, \ket{-_L}}$ for the $X$ memory.
In addition to the raw $\{0,1,2\}$ measurement values, we provide the neural network with calculated detector outcomes and logical operator values.
These additional features help the model learn from structured correlations in the data.
Detector values are computed as binary parities (0 or 1) over specified stabilizer regions; if a measurement gives a loss (2), it is assigned a value of 0 when computing detector parities.
Logical operator values are calculated along each row or column, depending on the basis.
We use a hidden layer size of 1024, the Adam optimizer with an initial learning rate of $10^{-3}$, and a weight decay of $10^{-2}$.
The learning rate is decreased by a factor of $0.3$ if the validation loss does not improve for 10 epochs.
Training is performed independently for 10 total experimental configurations: two with code distance $d=5$ (in the $Z$ and $X$ bases) and eight with $d=3$ (covering four spatial quadrants in both bases).
In the pre-training phase, each model is trained on 200 million simulated shots and validated on 20 million simulated shots.
We find that decoder performance is largely robust to small perturbations in the error model, and thus precise tuning of simulation parameters is not necessary.
For a batch size of $\sim10^4$ shots, the inference time per shot is $0.33\,\mu$s on a GPU (NVIDIA-A100).

\textit{Ensembling ---}
To account for training variability and enhance robustness, we repeat the full training procedure with 10 different random seeds, resulting in 10 independently trained models per experiment. 
These are ensembled together by computing the geometric mean of their output probabilities. 
The resulting ensembled ML decoder achieves a logical error per round (LEPR) of 0.78(4)\% for $d=5$ and 1.37(3)\% for $d=3$.

\textit{Fine-tuning ---}
To improve decoding performance, we fine-tune each pre-trained decoder on experimental data taken from designated training sets (independent of the final dataset).
For the $d=5$ decoders, we fine-tune on approximately 37000 shots per basis. For the $d=3$ decoders, we use approximately 2500 shots per basis, per quadrant. 
The neural network architecture remains unchanged, and fine-tuning is performed using the Adam optimizer with a learning rate of $10^{-3}$ and a weight decay of $8 \times 10^{-2}$.
The resulting ensemble of fine-tuned ML decoder achieves a LEPR of 0.71(4)\% for $d=5$ and 1.33(4)\% for $d=3$.

\noindent{\it Hybrid ---} 
When comparing the MLE and ML decoders, we find they do not predict the same logical state on all shots and, in particular, differ on shots where one of the decoders has low confidence in its prediction. 
To further enhance performance, we therefore construct a hybrid decoder that combines the output confidences of the ensembled ML decoder with those from the delayed-erasure MLE decoder, where the MLE confidence is derived from comparing the probabilities of the most-likely error and the most-likely error which gives the opposite logical outcome.
The final prediction is given by the weighted geometric mean of the two confidence values with weights of 0.4 and 1 for the MLE and ML, respectively.
This results in a final value for the reported LEPR of 0.62(3)\% for $d=5$ and 1.33(4)\% for $d=3$, which corresponds to the ML with loss decoder reported in Fig.~2. \\

\noindent{\it MLE decoder - lattice surgery}.
In the lattice surgery experiment (Figs.~3c,d), we perform a joint \(ZZ\) measurement using additional stabilizer checks along the common vertical edge between the two surface codes (``seam''). We start with both codes prepared in $\ket{+_{L}}$ and perform two rounds of stabilizer checks on the effective $d\times2d$ surface code lattice, measuring both the codes' stabilizer checks and the new seam checks in each round. To measure the \(ZZ\) parity of the resulting logical Bell state, with the delayed-erasure MLE decoder~\cite{Baranes2025,Cain2024a}, we use two decoding procedures. First, we use only the ancilla measurements to obtain the result of the lattice surgery $Z_1^LZ_2^L$ measurement given by the product of the seam \(Z\) stabilizers. Second, we obtain $Z_1^LZ_2^L$ directly from the data qubit measurements, using the prior ancilla measurements in decoding. This measures the $ZZ$ parity of the logical Bell state obtained using lattice surgery.
Note that the seam checks from the final data qubit measurement are not included. A shot is counted as an error, $Z^1_LZ_L^2 = -1$, if these two decoding procedures disagree.  

To obtain the \(XX\) Bell state parity, we measure all data qubits in the $X$ basis and decode the joint $X^{1}_{L}X^{2}_{L}$ operator spanning both codes. The final logical error probability is given by the mean of the \(XX\) and \(ZZ\) parities. \\

\noindent{\it ML decoder - deep circuits.} To decode the 1D and 2D cluster states of logical [[7,1,3]] and [[16,6,4]] codes in Fig.~6, we employ a convolutional neural network (CNN). 
Since error correlations in the cluster state do not propagate beyond two CZ gates (see Fig.~6e), a convolutional window of size 3 is sufficient to capture the relevant correlations.
The decoder architecture comprises three components: an encoder, a convolutional block, and a readout module.
Both the encoder and readout are constructed from linear layers interleaved with GELU activations, with $\texttt{hidden\_size}=128$.

\begin{lstlisting}[language=Python, numbers=none]
encode = nn.Sequential(
    nn.Linear(input_size, 1024),
    nn.GELU(),
    nn.Linear(1024, 512),
    nn.GELU(),
    nn.Linear(512, 256),
    nn.GELU(),
    nn.Linear(256, hidden_size)
)
\end{lstlisting}

\begin{lstlisting}[language=Python, numbers=none]
readout = nn.Sequential(
    nn.Linear(hidden_size*8, 512),
    nn.GELU(),
    nn.Linear(512, 256),
    nn.GELU(),
    nn.Linear(256, 128),
    nn.GELU(),
    nn.Linear(128, out_size)
)
\end{lstlisting}

The convolutional block applied between the encoder and readout modules is defined as:

\begin{lstlisting}[language=Python, numbers=none]
conv = nn.Sequential(
    nn.Conv2d(hidden_size, hidden_size*2,
              kernel_size=3, padding='same'),
    nn.GELU(),
    nn.BatchNorm2d(hidden_size*2),
    nn.Conv2d(hidden_size*2, hidden_size*4,
              kernel_size=3, padding='same'),
    nn.GELU(),
    nn.BatchNorm2d(hidden_size*4),
    nn.Conv2d(hidden_size*4, hidden_size*8,
              kernel_size=3, padding='same'),
    nn.GELU(),
    nn.BatchNorm2d(hidden_size*8),
)
\end{lstlisting}

For 1D cluster state decoders, we replace the 2D convolutions with 1D convolutions and omit the batch normalization layers.

Training is performed using circuit-level simulations.
The decoder is tasked with inferring the signs of the logical cluster state stabilizers, which are of the form $X$ on a given qubit and $Z$ on its neighbors.
By performing measurements in alternating $X$ and $Z$ bases, half of the stabilizers can be reconstructed.
The remaining stabilizers are recovered by repeating the experiment with the measurement bases swapped.
The decoder predicts the stabilizer signs by inferring the initial state of the qubits measured in the $X$ basis.
All logical qubits are initialized in the $\ket{+_L}$ state, and software logical $Z$ flips are applied with probability $1/2$ to those measured in the $X$ basis, to generate a balanced training set.

The decoder input includes the raw measurement outcomes (0, 1, or loss), detector values computed from the measurements, and the raw logical operator values, similar to the input format used in the surface code decoder.
We train four distinct decoders: one for each combination of code type ([[7,1,3]], [[16,6,4]]) and cluster state geometry (1D, 2D).
Each model is trained on over 100 million simulated shots.

For further details on this decoder architecture, and on ML-based decoders for general quantum algorithms, see Ref.~\cite{BonillaAtaides2025}.\\

\small \noindent\textbf{Benchmarking surface code performance} \\

\noindent{\it NZNZ stabilizer gate pattern}. Here we describe the effective distance, defined as the minimum number of physical errors required to create a logical error, of $d$ rounds of repeated syndrome extraction using alternating ``N'' or ``Z'' movement patterns (ED Fig.~\ref{fig:ED_ExtraSurface}d).
By alternating gate orderings, the effective distance is close to the optimal value. To see this, note that without alternating orderings, the effective code distance in the rotated surface code is reduced by a factor of two due to hook errors~\cite{Dennis2002a} (from a theoretical perspective - see discussion at end).
A hook error is a physical error on the ancilla qubit halfway through the stabilizer measurement cycle that propagates onto two data qubits oriented parallel to the corresponding logical operator (e.g., $X_L$ for a physical $X$ error).
One of these propagated data qubit errors is immediately detected, while the other is detected in the following round by the next-nearest stabilizer along the direction of error propagation.
As a result, if the same gate ordering is used for each round of stabilizer measurements, a sequence of $\lceil\frac{d+1}{4}\rceil$ hook errors, one occurring in each round along the direction of error propagation, can generate a logical error upon correction. 
This issue is circumvented by alternating gate orderings between rounds, as only every other round has the unfavorable propagation.
In this case, $\frac{d-1}{2}$ physical errors on consecutive rounds are needed to generate a logical error.

In our experiments in particular, we choose an ordering of $N~Z~Z_r~N_r$, where $r$ represents performing the reverse ordering (see ED Fig.~\ref{fig:ED_ExtraSurface}d). In addition to such structuring helping preserve fault-tolerance against hook errors, we also note that the dominance of $Z$-type errors means that most errors do not lead to propagated errors between the middle two CZ gate layers. Due to these reasons, in simulations we do not observe having spatially alternating ``N'' and ``Z'' patterns helps performance (not plotted). \\

\noindent{\it Simulations.} 
We perform simulations using the Stim simulation package~\cite{Gidney2021a}.
We sample both Pauli errors and qubit losses. 
Pauli errors are generated using Stim’s sampling routines, based on circuit-level noise models. 
Qubit losses are sampled according to the loss probabilities associated with each instruction, and when a loss occurs, subsequent gates acting on the lost qubit are removed to reflect the absence of the qubit.
The simulations detect $\{0,1,\text{loss}\}$ during qubit readout, like in our experiments. For each set of physical parameters, we estimate the logical error rate via Monte Carlo sampling. Logical errors are declared when the decoder’s prediction for the logical observable differs from the true value. See Supplementary Information for details of the noise model, as well as the discussion below.\\

\noindent{\it Analysis of below-threshold performance for deep circuits.} 
In Figure 2 we perform four rounds of repeated QEC as a benchmark. 
However, increasing the circuit depth can affect the threshold in various ways, depending on the particular circuit. 
ED Fig.~\ref{fig:ED_LEPR}a shows how the LEPR ratio $r$ changes for a single logical qubit as we increase the number of QEC rounds using a theory error model, showing a roughly 17\% decrease in $r$ from 4 rounds to 20. 
Similarly, ED Fig.~\ref{fig:ED_LEPR}b plots the same quantity for a single logical qubit with an approximate experimental error model, showing an analogous 9\% decrease in $r$ from 4 rounds to 50. 
Further, by interspersing 1 transversal gate every 1 QEC round under an approximate experimental error model, we find the ratio $r$ changes by 2\% at 25 QEC rounds. 
Similarly, prior work with a theory error model has shown that the threshold can change by $\approx 10\%$ with 1 gate per QEC round~\cite{Cain2024a}.
These simulations indicate that the benchmark studied in Fig.~2 is representative, but depending on context, can be different on the scale of $\approx 15\%$ for deep circuits. 
We note, however, that in transversal architectures, the prevalence of logical gate teleportations (e.g., in magic state distillation and angle synthesis) makes it such that there are typically only several stabilizer measurement rounds before transversal measurement.

Our benchmark results are comparable to those in Ref.~\cite{Acharya2024}. For instance, although a one-to-one comparison is not direct due to the presence of loss information, using the supercheck metric reveals a 9.04\% mean detector error, comparable to the 8.5-8.7\% mean detector error in Ref.~\cite{Acharya2024}. \\

\noindent{\it Error budget and path to 10x below threshold.} To get to algorithmically relevant error rates of $\sim 10^{-10}$ \cite{Gidney2021, Beverland2022}, a factor of 5-10x below threshold can achieve the required errors with several hundred qubits in a code block \cite{Fowler2012}. Our performance is captured by the error budget in Fig.~2f, which we now describe in further detail. 

We first list our single-qubit errors and their possible improvements:

\begin{itemize}
    \item Local single-qubit gates are approximately 99.9\% fidelity, arising from a 0.05\% scattering error and residual miscalibrations.
    Increasing Raman detuning to 2.5 THz will further reduce scattering errors and miscalibrations from the Raman differential light shift, and improving calibration routines can thereby achieve 99.99\% fidelity. 
    \item Our coherence time in 852-nm traps is approximately 1-2 seconds depending on the dynamical decoupling sequence applied. Comparable systems have achieved coherence times of 12.6 seconds with further tweezer detunings \cite{Manetsch2024}.
    \item 
    We experience a total loss from movement of roughly 1\% on the ancilla atoms, arising from transfers and moves between and within the zones.
    We have previously observed performance in Ref.~\cite{Bluvstein2023} with transfer-limited loss that would correspond to 0.2\% movement loss here, which we speculate arises in the present work from using too high of an AOD RF power. 
    Our repeated QEC sequence also experienced 0.6\% background loss from vacuum that can be readily reduced to $<$0.01\% using improved vacuum lifetime and a shorter sequence (e.g. $\sim$4 ms cycle times in Fig.~5b).
    \item Our lattice readout currently is operating with a loss rate of 0.3\% and a 99.5\% bit-flip error rate. Although a new technique, similar methods in purely-lattice systems have achieved fidelities of 99.94\% \cite{Wu2019}. 
\end{itemize}

In order to improve two-qubit gate performance, an example approach can be:

\begin{itemize}
    \item Improve system stability, homogeneity and fast, automated calibration. Although we achieve CZ fidelities of 99.6\%, drifts since the last calibration (several days in the context of the surface code benchmarking) often contribute $0.05-0.1\%$ error during final data taking.
    \item Use the smooth-amplitude gate, higher magnetic fields, or the $6P_{1/2}$ intermediate state to suppress coupling to the adjacent $m_j = +1/2$ state, reducing the error from $\approx 0.06-0.15\%$ to near-zero.
    \item Increase both 420-nm and 1013-nm Rydberg laser power by a factor of 4x. This can allow to simultaneously (numbers are from simulation, see Ref.~\cite{Evered2023a}):
    \begin{itemize}
        \item Increase Rydberg detuning from 4.8 GHz in the present work to 9.6 GHz, reducing scattering error from 0.094\% to 0.052\%.
        \item Decrease gate time from 270 ns to 135 ns, reducing Rydberg $T_1$ error from 0.113\% to 0.057\%, and reducing dephasing error from 0.134\% to 0.034\%. 
    \end{itemize}
    
\end{itemize}

These can reduce the two-qubit gate error from roughly 0.5\% to 0.15\% through simple system improvements. The AOM pulse profile should be compensated for realizing these gate times, and the 1013-nm beam's fractional inhomogeneity will need to be improved by the corresponding increase in power.

Altogether, by reducing single-qubit gate errors by a factor of 5x and improving two-qubit errors from 0.5\% to 0.15\% through the improvements listed, operation at roughly 8x below threshold would be achieved. The two-order-of-magnitude increase in cycle rate demonstrated in Fig.~5b will be instrumental to enabling these improvements. These estimates highlight that straightforward improvements can lend the performance required for large-scale computation. We also emphasize that such performance will need to be tested and optimized in deep-circuit settings.\\

\small \noindent\textbf{Processor clock speed} \\

Future operation will eventually be impacted by the speed of operations, once algorithms with, for example, trillions of operations need to be realized \cite{Gidney2021, Beverland2022}. In the present work, we do not optimize for clock speed, and often choose slower speeds for our components so that they can function reliably without detailed characterization on existing infrastructure. However, we here report multiple measurements of our circuit durations.

In the repeated surface code experiments in Figure~2, each QEC round was 4.45 ms. 
This originated from 0.47 ms time between gates, and a total of 2.57 ms from moving the ancilla atoms to the storage zone and bringing in the next group to the entangling zone. In the transversal CNOT experiments in Figure~3, we fix the overall circuit duration (independent of number of CNOTs) at 17.7 ms, corresponding to the time of the longest circuit of 27 total transversal CNOTs. This corresponds to 0.655 ms per transversal CNOT on average. In the deep circuit experiments in Figure~6, our cycle rate was bottlenecked through the use of desktop computers for all data processing for the mid-circuit image analysis and rearrangement, and so we did not attempt to reduce any times. For this reason each logical teleportation layer was 41.9 ms.

In the repeated Rabi calibration in Fig.~5b, we did optimize for speed and achieved a cycle time of 4 ms. Although the imaging here was global as a demonstration of fast calibration, we expect such speeds can also be achieved in a zoned manner.

We thereby expect that, with optimization for speed in the deep-circuit context and various simple improvements, one should be able to achieve a logical teleportation cycle with a cycle time comparable to the 4-ms repeated Rabi calibration. We emphasize that in a planar architecture, such a logical teleportation step involves multiple logical gates and can require multiple hundreds of QEC rounds for large-distance codes, e.g. 200-300 stabilizer measurement rounds. As such, we estimate the present methods are slower by a factor of $\sim 10-20$ relative to a conventional planar architecture with $1 ~\mu s$ speed per stabilizer measurement cycle \cite{Gidney2021, Acharya2024}.\\

\small \noindent\textbf{Physical entropy removal} \\

\noindent{\it Types of entropy.}
QEC enables removing entropy from the physical qubits, and this entropy can take on many different forms. As discussed above, error correction such as stabilizer measurement serves the role of converting generic quantum errors, such as coherent ones, into incoherent bit- and phase-flip errors. 
Detecting and tracking these errors further removes entropy from the system. Finally, physical systems such as atoms have entropy in other degrees of freedom such as loss, leakage, or atom heating. We would like to design our QEC strategy to remove all of these entropy types.
\\

\noindent{\it Overview of entropy removal methods}. Ancilla-based stabilizer extraction, as used in Figs.~1-3, is one form of entropy removal, in which stabilizer information is mapped onto the ancilla and then the ancilla is measured. Shor-style syndrome extraction operates via entangling ancillas into a GHZ state and extracting the stabilizer in a single step \cite{Shor1996}, Steane-style syndrome extraction operates via creating an ancilla logical qubit and extracting stabilizers via a transversal CNOT \cite{Steane1997}, and measurement based quantum computing (MBQC)-style syndrome extraction operates via sequential entanglement with adjacent layers \cite{Raussendorf2001, Briegel2009, Sahay2023a}. Leveraging teleportation native to the algorithm is another related method of entropy removal without ever `directly' correcting the initial logical qubit block after it was used in computation. While these methods all vary in their specific implementations, their core mechanism of entropy removal is similar. These methods can be used interchangeably depending on specific practical considerations, such as those discussed in the next section. \\

\noindent{\it Use of teleportation for ensuring error removal.} Logical teleportations are a method to ensure an architecture natively removes physical errors such as bit- and phase-flips, but also physical errors such as loss, leakage, and heating \cite{Raussendorf2001}. In particular, by teleporting a logical qubit from one block to another, the logical information propagates but the physical errors---both Pauli-type and other complex errors---are all left behind. This method ensures errors of all types are removed. Due to transversal gates leading to only $O(1)$ QEC rounds per logic gate, algorithms can be composed of a high density of logical gate teleportations. This highlights that, as shown in Figure~6, teleportation can perform logical operations while natively removing all such errors without additional overhead.

An alternative method for removing physical errors is teleportation at the physical level -- specifically, by swapping quantum information between a physical data qubit and a physical ancilla.
This approach underlies various implementations of leakage reduction units \cite{Sahay2023a,Miao2023, Chow2024a,Yu2025,Baranes2025}.
However, it necessitates pairing each data qubit with a dedicated ancilla, which can present challenges.
For example, in high-rate quantum LDPC codes encoding many logical qubits, such one-to-one pairing can become increasingly impractical.
In general, the number of unpaired data qubits in each round of error correction is lower bounded by $k=$ \# data qubits - \# independent checks, where $k$ is the number of encoded qubits.
For example, in hypergraph product codes constructed from ($u$, $v$)-biregular expanders -- bipartite graphs in which checks have degree $u$ and bits have degree $v$ -- the compact rearrangement scheme of Ref.~\cite{Xu2024} implies that there will be $O((v - u)d)$ unpaired data qubits per error-correction cycle, where $d$ is the distance of the code.
In contrast, logical-level teleportation is directly accessible in all CSS codes (as they all have a transversal CNOT), as demonstrated in the high-rate [[16,6,4]] code in Figure~6. Such analysis highlights that leveraging the transversal teleportations native to an algorithm lends to a robust, low-overhead procedure that ensures all physical errors are removed independent of the specific code.\\

\noindent{\it Feedforward in universal processing.} Once bit-or phase-flip errors have been detected, a natural question is if they need to be physically corrected in-hardware in order to return back to a configuration with all stabilizers = +1. For conventional computation based on transversal (or planar) Clifford gates, stabilizer measurements, and universality achieved via teleportation of $\ket{T_L}$ states (realized via physical Clifford gates), one does not have to apply such physical qubit corrections. This can be most directly seen by the fact that universal computation on the logical-qubit level is realized via physical Clifford gates \cite{Knill2005, Fowler2012}, and so the Pauli corrections can be deterministically tracked in-software as a Pauli frame update without additional overhead on the decoding.

When realizing transversal non-Cliffords, such as the transversal $T$ gate in the [[15,1,3]] Reed-Muller code \cite{Bravyi2005}, X Pauli corrections do not commute through, and so in such a case the stabilizers do need to be returned to a deterministic $+1$ eigenvalue. However, in the results here, for example, we realize deterministic initialization of the Reed-Muller code with +1 eigenvalues as a method of ensuring constant entropy operation, and in such a case even here mid-circuit correction of individual physical qubits is not required.

In both of these settings, feedforward is indeed required, but only on the logical-qubit level (feedforward $S$ for $T$ teleportation, and feedforward $X$ for $H$ teleportation). We implemented such a logical feedforward in Fig.~4 of Ref.~\cite{Bluvstein2023} where feedforward logical $S$ gates were realized to entangle two qubits that did not directly interact.
\\

\noindent{\it Transversal logic with O(1) stabilizer measurements per gate}. Since in the transversal setting, the role of stabilizer measurements is simply to remove entropy, we do not require the conventional $d$ rounds of stabilizer measurement per logic gate, as shown in Figure 3 of the main text. We note that these techniques directly apply for universal computation. Concretely, universal computation is implemented by a transversal Clifford circuit, where $T$ gates are realized via a transversal teleportation circuit with $\ket{T_L}$ inputs that have already been prepared fault-tolerantly. It has been shown that this universal processing can proceed with $O(1)$ stabilizer measurements per transversal gate, and that the decoding can also be done efficiently with a decoding complexity that can be in fact even less than the conventional lattice surgery setting \cite{Delfosse2023, Cain2024a, Zhou2024, Cain2025, Serra-Peralta2025, Turner2025}. As such, our experimental results directly apply to universal computation, under the assumption that the $\ket{T_L}$ inputs are prepared to high quality.
\\

\small \noindent\textbf{Methods of universality} \\

\noindent{\it Universality, transversal gates, and Eastin-Knill.} Universality means that any unitary can be closely approximated by using sequences of gates from a universal gate set \cite{Kitaev1997}. An example universal gate set is $\{H, T, CNOT \}$. 2D topological codes can have a discrete gate set of $\{H,S,CNOT\}$, but cannot transversally implement the $T$ gate. 3D topological codes can in fact have a transversal $T$ gate \cite{Kubica2015}, and the [[15,1,3]] 3D color code in particular has a transversal gate set of $\{CZ,CCZ,CNOT,T\}$. The Eastin-Knill theorem forbids having a unitary transversal gate set which is universal \cite{Eastin2009}. This is expected, as this would, for example, allow realizing a transversal logical $\theta$ rotation by a sequence of transversal operations on the underlying physical qubits, and thereby could not be protected as it would be sensitive to small imperfections in the physical rotation.

Crucially, the Eastin-Knill theorem is easily circumvented simply by the introduction of logical measurement, which breaks unitarity and enables universality. This is directly achieved with 3D codes, as realizing a CZ gate between state $\ket{\psi_L}$ and $\ket{+_L}$, followed by logical measurement and feedforward, teleports a Hadamard gate directly onto $\ket{\psi_L}$. As such, X-basis preparation and X-basis measurements (guaranteed in all CSS codes), combined with transversal CZ gates, can be used to straightforwardly implement a universal gate set of $\{H,T,CNOT\}$ using fully transversal operations. This is the basis behind our implementation of universality in Figure~4.

We note that in many protocols, universality is directly generated via the measurement of a 3D code. Code switching is one example, where one switches between codes that have $T$ and $H$ transversal gates \cite{Anderson2014}. For example, we realize a code switching protocol in ED Figure \ref{fig:ED_Universality}e, where we teleport a logical T from a 3D [[15,1,3]] color code onto a 2D [[7,1,3]] color code. Such operations between codes of different dimensionality can often be realized, and here just involves entangling the 2D surface of the 3D pyramid with the 2D color code face. While teleportation onto the 2D color code now admits transversal $H$ gates, this is anyway already accomplished via the transversal measurement and feedforward from the 3D code. \\

\noindent{\it Connection to magic state distillation.} We note that the protocols studied here are similar to those underlying magic state distillation \cite{Bravyi2005, Campbell2012, Howard2017}. In the conventional 15-to-1 magic state distillation, 16 surface code logical qubits are entangled in a manner where the first surface code qubit is entangled with the logical qubit of a [[15,1,3]] code made out of surface codes. Subsequently, noisy $T$ gates with some error $p$ are realized on the surface codes via teleportation, which the outer [[15,1,3]] code distills into $\sim p^2$ with correction or $\sim p^3$ with postselection \cite{Litinski2019}. By measuring the Reed-Muller code, the resulting distilled $\ket{T}$ state is teleported onto the first surface code.

The protocol we study in Figure~4 is a more compact representation of the same magic state distillation circuit, but with replacing the inner surface codes with unencoded physical qubits. While in conventional distillation the $\ket{T}$ is teleported onto the surface code, we note that, for example, in small-angle synthesis with sequential $HTHT...$ gates, one does not even need to do the step of teleporting onto the surface code - one can simply leave the $\ket{T}$ encoded in the Reed-Muller code and then realize a transversal CZ gate between the two concatenated Reed-Muller codes. \\

\small \noindent\textbf{Physical resources for QEC} \\

In this work we study the relationships of many different physical resources, and how they are used in quantum error correction. We overview here some of our observations and discuss how these can be useful for developing future QEC protocols and architectures.\\

\noindent{\it Logical entanglement and physical entanglement.} In a transversal gate setting, logical entanglement can be generated using only physical entanglement between the code blocks. This is in contrast to lattice surgery, where entanglement within the blocks is necessary to mediate interaction between non-overlapping logical operators, and so one needs a robust entanglement. This is the origin of the sensitivity to measurement errors in the lattice surgery context, and the insensitivity to measurement errors in the transversal gate context. Logical entanglement within the code block also plays an interesting role. The [[16,6,4]] codes, for example, contain many logical qubits within the block, which can be entangled, but only with a sufficient degree of physical entanglement present (discussion below).

Motivated by these observations, one way to re-frame efficient encodings is to find methods that generate the target logical entanglement with the minimum amount of physical entanglement. To this end, we first prove that the amount of logical entanglement - even generated with techniques such as permutation gates - cannot exceed the amount of physical entanglement.

Operator entanglement quantifies the maximum entanglement a gate can produce on separable inputs. For any gate acting on $k$ qubits, the operator entanglement is bounded above by $\lfloor k/2 \rfloor$. Thus, for a quantum code with parameters $[[n, k, d]]$, the logical operator entanglement satisfies $S_{\rm LO} \leq \lfloor k/2 \rfloor$.
To lower bound the physical entanglement entropy, note that logical operators cannot be supported on any set of $d-1$ or fewer physical qubits. Therefore, for any region $A$ with $|A| \leq d-1$, all logical codewords yield identical reduced density matrices on $A$ (not necessarily maximally mixed). For the maximally mixed logical state $\rho_L$, the reduced density matrix on $A$ has rank $2^{\min(k, d-1)}$, implying $S_{\rm PS} \geq \min(k, d-1) $. If $k \leq d-1$, then $S_{\rm PS} \geq k$, so $S_{\rm PS} \geq S_{\rm LO}$ always holds. If $k > d-1$, then $S_{\rm PS} \geq d-1$. Here, $S_{\rm LO} \leq S_{\rm PS}$ as long as $d-1 \geq \lfloor k/2 \rfloor$, i.e., $k < 2d$.

Thus, for any $[[n, k, d]]$ code with $k < 2d$, the logical operator entanglement cannot exceed the physical entanglement available in any region of size $d-1$.

These observations can have applicability to finding efficient algorithm compilations with high-rate codes and transversal operations, both of which we observe here can reduce the amount of physical entanglement to realize the target logical entanglement. For instance, each transversal CNOT in the [[16,6,4]] code generates 16 physical CNOTs worth of entanglement and 6 logical CNOTs worth of entanglement, but realizing in-block permutation CNOTs can generate an additional 4$\times$2 logical CNOTs, totaling 14 logical CNOTs worth of entanglement, close to the bound of 16 physical CNOTs.
\\

\noindent{\it Physical entanglement and logical magic.} While physical entanglement is the underpinning of logical entanglement, it is also the underpinning of logical magic. In particular, we find here that states with logical magic require more in-block entanglement than states without any logical magic. This can be understood by the fact that, while logical Pauli states such as $\ket{+_L}$ are represented by operators $X_L =  X_1X_2X_3...$ (in CSS codes), which is a tensor product of physical operators, states such as $\ket{T_L}$ are represented by $\frac{1}{\sqrt{2}} (X_1X_2X_3... + Y_1Y_2Y_3...)$, involving a macroscopic superposition of operators spanning the code that is necessarily entangled \cite{Korbany2025,Wei2025}. Analogously, any physical product state that has deterministic $X$-type stabilizers must have zero expectation value for $Y_L$. The need for well-defined stabilizers in both bases is thereby another way to see that the code must be entangled. Similarly, CSS codes are constructed from two classical codes \cite{Calderbank1996,Shor1995, Shor1996, Steane1996}, and Pauli states are `classical' in that they store 1 bit of information in one of the two classical codes (and 0 bits in the other), whereas $\ket{T}$ states truly require both codes. 

Intriguingly, these observations suggest a potentially more fundamental mechanism of what algorithmic outputs do and don't need full protection. For example, consider making a remote entangled Bell pair. To probe its fidelity with $X_LX_L$ and $Z_L Z_L$ entanglement witnesses, then with correlated decoding methods one does not need a high degree of entanglement within the individual code blocks - just between the blocks. However, if one would instead like to perform an error-corrected Bell inequality test \cite{Bell1964}, to provide evidence that quantum mechanics is real, then now one has to measure in the $\ket{T}$ basis and requires the full entanglement within the block. It has been argued that so-called quantum contextuality, which arises from measurements in non-Pauli bases, is the core aspect of quantum mechanics that cannot be described by classical theories \cite{Spekkens2005}. Relatedly, theoretical work has shown a connection between contextuality and computational hardness \cite{Howard2014}, and in this work, we find that both of these are also linked to the minimum amount of entanglement required to perform the requisite error correction. Understanding the essence of these connections may hint to further avenues to reduce resource requirements for protecting the relevant algorithmic outputs. An experimental error-corrected Bell inequality test is shown in ED Fig.~\ref{fig:ED_Universality}f. \\

\textit{Logical gate fidelity and physical entropy.} With physical qubits, which are two-level systems, fidelity is a descriptive and accurate concept as noise can often be decomposed into either realizing the correct operation or the exact opposite (e.g., a bit-flip error). Conversely, logical qubits are many-level systems, and so this property does not hold. This fact is related to our observation in Fig.~3d that the error per logical operation is not constant as a function of the number of applied logical gates. Instead, there is a logical fidelity associated with the probability of decoding correctly \cite{Fowler2012}, which depends on the internal density of errors $p$. 

Theoretically, the per-step logical error from decoding scales approximately as $P_L \propto (p/p_{th})^{(d+1)/2}$ \cite{Fowler2012}. The results shown in Fig.~3d clearly indicate that quantifying logical gate performance should encapsulate $\{F_L(p_{\text{det}}),\Delta p_{\text{det}} \}$, which capture how the logical fidelity $F_L$ depends on the internal density of errors $p$, as well as the gate's increase to local error density $\Delta p$. We study this quantitatively in ED Fig.~\ref{fig:ED_Coherent}. \\

\small \noindent\textbf{Additional experiment and data analysis details} \\

\textit{Figure 1. } 
In Fig.~1d, we prepare either $\ket{+_L} = \ket{+}^{\otimes 25}$ or $\ket{0_L} = \ket{0}^{\otimes 25}$ and apply up to five rounds of stabilizer measurement followed by measurement in the X or Z basis, respectively. A global $Z(\theta)$ rotation is applied to the data qubits at every gate layer (20 time-steps in total). For fewer than five stabilizer measurement rounds, the relevant CZ gates are removed but single qubit rotations still applied. 1 QEC round has gates in the first round only, 2 QEC rounds has gates in the first and fourth rounds, and 5 QEC rounds has gates in all five rounds. The data is averaged over both initial states and uses MLE decoding with a 50\% acceptance fraction for visual clarity, as well as pre-selection on perfect initial qubit filling. The right plot uses an injected error of $\theta/2\pi = 0.016$ and additional error rates are shown in ED Fig.~\ref{fig:ED_Coherent}b. \\

\textit{Figure 2.} No postselection is used in the analysis of the surface code. Pre-selection of initial qubit rearrangement (standard in the literature) is used. 
Data in Figs.~2b-d are averaged over $\ket{+_L}$ and $\ket{0_L}$, and the distributions in Figs.~2e,g aggregrate the two bases.
Figs.~2b,c plot the detector error probability averaged over all rounds. Fig.~2b uses the same data set with shots binned according to data qubit loss.
The four metrics are (i) `bare', where loss is converted to qubit state 0 (ii) `detect loss', where projective measurements whose value is `loss' are not counted erroneously (iii) `supercheck', where stabilizers with a lost data qubit are formed into superchecks for all prior rounds, and (iv) `postselected', where detectors involving any lost atoms are ignored.
The plots show the mean error of all deterministic detectors (96 per basis). 
The supercheck error is calculated over all samples per round per basis, and the mean of these 8 values plotted. Superchecks paired to the boundary are removed from the averaging as these return no error by construction; if included, the mean error decreases from 9.0\% to 8.8\%. The contribution of each supercheck is normalised by the supercheck weight, e.g., a weight-6 supercheck contributes an error of 4/6 (to account for the greater amount of information in the check - e.g., multiplying checks even in absence of loss raises the detector error without reducing the amount of information). Without reweighting, the error probability increases to 9.6\%.  

In Figs.~2d,e, the logical error per round is calculated as $\mathrm{LEPR} = 0.5(1-(1-2p_L)^{1/r})$ where $p_L$ is the final logical error after $r=4$ rounds, same as the definition in Ref. \cite{Acharya2024}.
The $d=5$ dataset contains 9021 shots in the X basis and 5834 in the Z basis. The $d=3$ dataset contains 2523 shots for X and 2534 for Z (on average per quadrant).  To make a $d=3$ surface code in each of four possible quadrants, we only remove atoms and do not modify the circuit.
The specific circuit for the repeated stabilizer measurement is shown in ED \ref{fig:ED_ExtraSurface}c. Not shown are local Y$(\pi)$ and Y$(\pi/2)$ gates on the boundary ancillas (see Supplementary Information and Stim circuit).
Additionally, local detunings \cite{Manovitz2025} are applied to the lowest row of gate sites (where there are only isolated ancilla qubits) to mitigate inhomogeneity in the 1013-nm lightshift during entangling gates.

In Fig.~2f, the error budget shows the contributions to the detector error (with loss detection) and is obtained by removing error sources individually from the simulation error model. We obtain a similar error model breakdown by simulating the relative contribution to the logical error. Fig.~2g plots the detector distribution with loss detection. ED Fig.~\ref{fig:ED_ExtraSurface}f compares the bare detector error to simulation. 

See Supplementary Information for the error model including quantitative error budget and pseudocode for simulation, an animation showing the moves realized experimentally, and an annotated version of the raw experimental command strings used to realize the circuit. See Ref.~\cite{Bluvstein2025a} for all raw experimental shots, the analysis notebook, and trained machine learning decoders. \\

\textit{Figure 3.} All data uses MLE decoding and is pre-selected on perfect initial qubit filling. In Figs.~3c,d, we prepare logical Bell states using either transversal gates or lattice surgery, and measure the mean error in the resulting $XX$ and $ZZ$ parities. The error per logical operation is defined as $\varepsilon = 0.5 (1-(1-2p_L)^{1/3N})$ for $N$ transversal gates per round and Bell state infidelity $p_L$, and $\varepsilon = p_L$ for the lattice surgery logical product measurement.
In Fig.~3c, the transversal CNOT is shown for 3 CNOTs per QEC round, and the injected measurement error is applied in-software with probability $p$ independently to every ancilla qubit. An acceptance fraction of 1 is used for the transversal CNOT plots unless otherwise stated. In Fig.~3d, the lattice surgery point uses error detection on the middle three ancillas each having the same value in both rounds of stabilizer measurement, to compensate for having fewer than $d$ rounds of repeated syndrome measurement for this result. We find in numerical simulations using our experimental error model that the optimal number of QEC rounds for this circuit is approximately 3 (as opposed to 5), and that by using error detection with 2 rounds we recover a similar performance to this optimal value found in numerics.
\\

\textit{Figure 4.}  
To modify the stabilizer signs in Fig.~4a, local $\pi$ pulses are applied at the end of the encoding circuit. ``Negative'' stabilizers corresponds to flipped qubits on the four corners of the Reed-Muller tetrahedron.
We use a lookup table for decoding and plot all three 3D color code curves with an acceptance fraction of 46\% and the 2D color code with 74\%, corresponding to a rescaling by the number of physical qubits in the code. For postselection, the shots are ordered by the weight of the detected error. The curves are further normalized to highlight key features, with unnormalized data shown in ED Figure \ref{fig:ED_Universality}a.
Fig.~4c uses error detection and plots the angles for $\leq N$ T gates. All plots are postselected on no loss and perfect initial qubit filling.\\

\textit{Figure 5.} 
In Fig.~5c, we study the atom temperature and loss as a function of cycle using the circuit for state preparation of Steane codes (Fig.~6b) with only entangling gates removed. In the fifth cycle, we turn off all imaging and cooling light. For comparison, the same measurements are repeated with conventional 3D PGC imaging and cooling in place of the local techniques. To extract the atom temperature shown in the upper panel, we use a drop-recapture measurement after $N$ cycles and fit the resulting loss to a Monte-Carlo simulation \cite{Tuchendler2008}. Shaded regions indicate the range of fitted temperatures due to uncertainty in trap parameters.
\\

\textit{Figure 6.}
The [[7,1,3]] and [[16,6,4]] codes in Fig.~6, as well as the [[15,1,3]] in Fig.~4, are members of the family of quantum Reed-Muller codes based on the hypercube encoding circuit illustrated in ED Fig.~10a (see also Supplementary Videos) \cite{Gong2024}. For each code, a different pattern of local Y($\pi$/2) pulses is applied while the entangling gate structure is the same; for the 2D [[7,1,3]] code, the fourth layer of gates is turned off. In Fig.~6b, groups of sixteen independent [[7,1,3]] codes are prepared in parallel in each time layer, repeated for 27 layers. The stabilizer error probability as a function of layer is plotted for no loss in the code block.

To characterize the propagation of physical and logical information in deep circuits, we further entangle the codes into 1D and 2D cluster states. Starting with two groups of logical qubits, group A and group B in Fig.~6a, these are entangled to form the first two time layers of the cluster state. Due to the local entanglement structure of a cluster state, group A undergoes no more entangling gates and is idle up until its measurement (in the appropriate basis), and thereby the measurement can be performed and the same physical qubits re-used to form the third layer of the cluster state (in typical MBQC fashion). Group B can then be measured and re-used to form the fourth layer of the cluster state, and so on. This alternating structure is typical in MBQC using cluster states \cite{Raussendorf2001}.

The physical correlations in Figs.~6c,d,g are calculated as the covariance between errors (stabilizer = -1) on the same stabilizer between codes at different coordinates in the cluster state. 
The covariance is then averaged across all co-propagating cluster states and the different stabilizers (three for [[7,1,3]] and five for [[16,6,4]]). 
The logical correlations in Figs.~6c,g are calculated as the appropriate product of cluster state stabilizers between the two target coordinates (cluster states have stabilizers corresponding to $X_i \Pi_{j} Z_j$ where $i$ is a specific site and $j$ is its neighbors). For example, we define $\langle Z_0 Z_4 \rangle \equiv \langle (Z_0X_1Z_2)\cdot(Z_2X_3Z_4) \rangle$. The single-qubit expectation values $\langle Z_i \rangle$ are calculated using a lookup table decoder for [[7,1,3]] and raw values for [[16,6,4]].
Due to the underlying assumption of time- and space- invariance, i.e., that correlations depend only on relative coordinates, we truncate time layers where the reservoir begins to be depleted and this assumption breaks down. This corresponds to 16 layers for Fig.~6c, 13 layers for Fig.~6d (correlations plot only), and 12 layers for Fig.~6g. This has only a small effect on the measured logical correlations but otherwise leads to a longer-tail of physical correlations because atoms are not properly refilled once the reservoir begins depleting.

All logical operators in Figure~6 are decoded with machine learning which directly predicts the cluster state stabilizers.
The 2D cluster state stabilizers in Fig.~6d use a global acceptance fraction of 0.24\%. 
In Figs.~6c,g we instead use a global confidence threshold for each curve, which is then converted to a mean acceptance fraction. In this way, each curve corresponds to a constant effective error rate for the logical operator independent of its weight, resulting in a reduced acceptance fraction for higher weight operators. The confidence for products of the weight-3 logical stabilizers is given as the geometric mean of the constituent confidences. 
Fig.~6g uses a mean acceptance fraction of 3.4\% (same data for both curves).
The 2D [[16,6,4]] cluster state in Fig.~6i also uses the confidence-based postselection, where the confidence per cluster state stabilizer is the geometric mean of the six decoded co-propagating 2D cluster states. On top of this decoding postselection, the logical stabilizer expectation value is shown as a function of the minimum number of co-propagating operators, N, with the same measurement outcome. We take the mean of all combinations of choosing N out of 6 such operators.

The permutation CNOT in Fig.~6g is applied in software and its effect here is to increase the weight of the operator connecting coordinates $t_i$ and $t_j$, labeled as an effective separation $i-j$. Following the definitions in Ref. \cite{Reichardt2024a}, two permutation CNOTs (swapping a pair of rows and a pair of columns) convert the cluster state stabilizers supported on logical qubits 3 to 6 from four weight-3 to one weight-3, two weight-6 and one weight-12 operator. 

See Supplementary Information for an annotated version of the raw experimental command strings used to realize the circuit.
\\

\noindent\textbf{Data Availability}\\
The data that supports the findings of this study are available from the corresponding author on request. The raw data for the surface code repeated QEC is available online in Ref.~\cite{Bluvstein2025a}.\\

\noindent\textbf{Acknowledgments} \\
We thank H.~Levine for technical and scientific discussions as well as detailed feedback on the manuscript, N.-C.~Chiu, J.~Guo, M.~Abobeih, P.~Stroganov, L. M. Peters, and T. T. Wang for key discussions about continuous operations, and Dan Kleppner and Ray Laflamme for building scientific foundations and communities in AMO and quantum information that this work builds upon. We also thank S. Cantu, C.-F.~Chen, S. Choi, J. Cong, M. Devoret, S. Ebadi, M. Endres, H.-Y. Huang, P. N. Jepsen, S. Kolkowitz, A. Kubica, A. Lukin, J. Preskill, H.~Pichler, H.~Putterman, J. Robinson, P. S. Rodriguez, D. Tan, N. Uğur Köylüoğlu, E. Qiu, T. Šumarac, S. Tsesses, Q. Xu, and P. Zoller for helpful discussions. 
We acknowledge financial support from the 
IARPA and the Army Research Office, under the Entangled Logical Qubits program (Cooperative Agreement Number W911NF-23-2-0219),
DARPA ONISQ program (grant number W911NF2010021) and MeasQuIT program (grant number HR0011-24-9-0359), the Center for Ultracold Atoms (an NSF Physics Frontier Center), the National Science Foundation (grant numbers PHY-2012023, CCF-2313084 and QLCI grant OMA-2120757) , the Army Research Office MURI (grant number W911NF-20-1-0082),  Wellcome Leap Quantum for Bio program, and QuEra Computing. 
D.B. acknowledges support from The Fannie and John Hertz Foundation. S.J.E. acknowledges support from the National Defense Science and Engineering Graduate (NDSEG) fellowship. G.B. acknowledges support from the MIT Patrons of Physics Fellows Society. A.G. acknowledges support from the IBM PhD Fellowship. J.P.B.A. and A.G. acknowledge support from the Unitary Foundation for developing the machine learning software used in the results.  T.M. acknowledges support from the Harvard Quantum Initiative Postdoctoral Fellowship in Science and Engineering. SFY and AG acknowledge the NSF through the HDR Q-IDEAS grant (OAC-2118310). S.M. acknowledges support from the Banting Postdoctoral Fellowship. C.K. acknowledges support from the NSF through a grant for the ITAMP at Harvard University. N.M. acknowledges support by the Department of Energy Computational Science Graduate Fellowship under award number DE-SC0021110. M.C. acknowledges support from Department of Energy Computational Science Graduate Fellowship under Award Number DE-SC0020347. The commercial equipment used in this work does not reflect endorsement by NIST.
\\

\noindent\textbf{Author contributions} D.B., A.A.G., S.H.L., S.J.E, T.M., M.X., and M.K., contributed to the building of the experimental setup, performed the measurements, and analyzed the data. J.P.B.A., G.B., A.G. and M.C. developed and implemented the decoding infrastructure. S.M., C.K., N.M., and H.Z. performed theoretical analysis.  E.C.T., L.M.S., and S.H. contributed to atomic replenishment methods. All work was supervised by M.J.G., S.F.Y., M.G., V.V., M.C., and M.D.L. All authors contributed to the project vision and understanding about the physics of fault-tolerance, discussed the results, and contributed to the manuscript.
\\

\noindent\textbf{Competing interests:} M.G., V.V., and M.D.L. are co-founders, M.G. V.V., M.D.L, and H.Z. are shareholders, S.F.Y.’s spouse is a co-founder and shareholder, V.V. is Chief Technology Officer, M.D.L. is Chief Scientist, and H.Z. is an employee of QuEra Computing.\\ 

\noindent\textbf{Correspondence and requests for materials} should be addressed to D.B. and M.D.L.\\


\setcounter{figure}{0}
\newcounter{EDfig}
\renewcommand{\figurename}{Extended Data Fig.}

\begin{figure*}
\includegraphics[width=2\columnwidth]{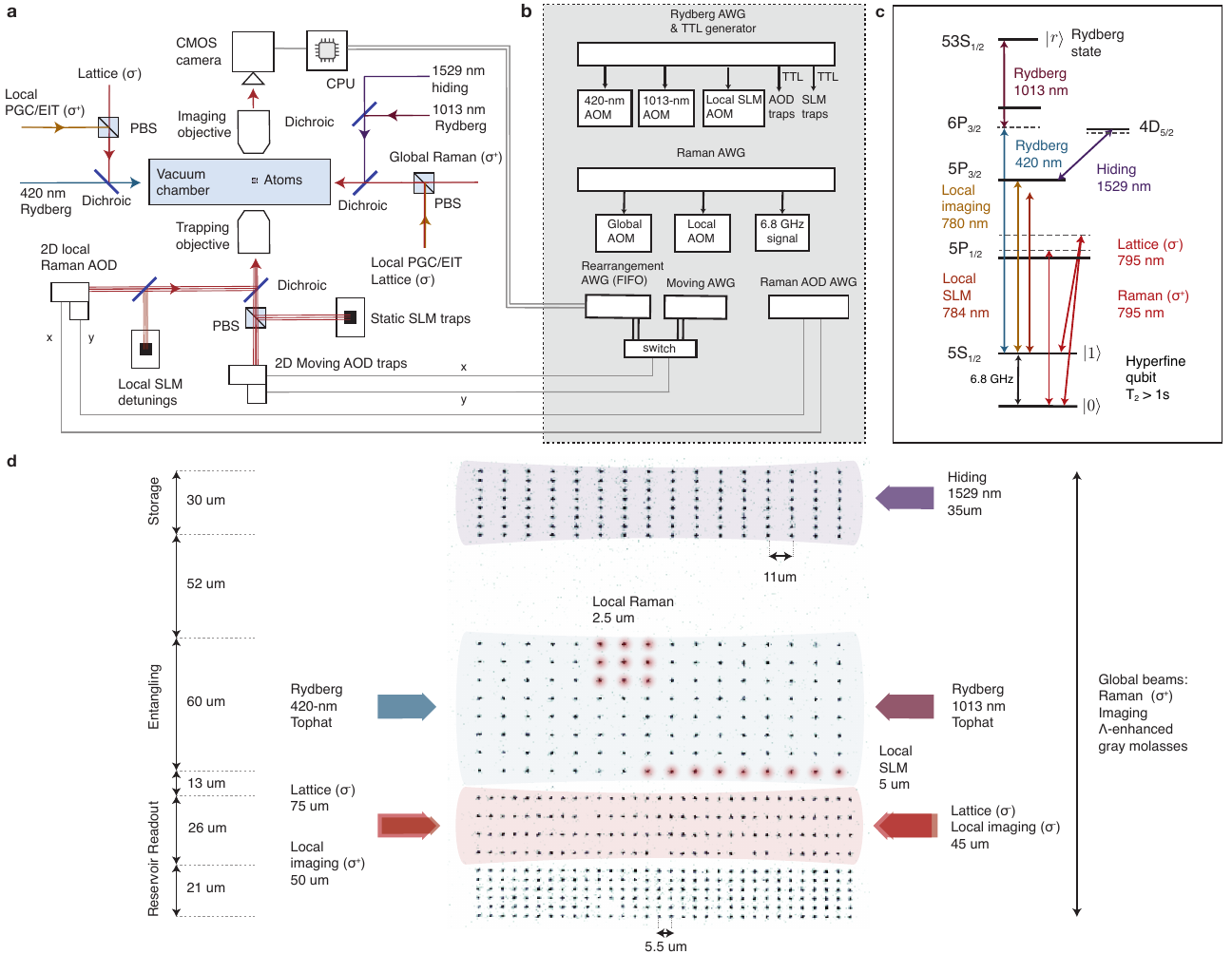}
\caption{\textbf{Neutral-atom quantum computing architecture.} \textbf{a,} Experimental layout illustrating key optical tools, similar to Ref. \cite{Bluvstein2023} with the addition of beams for local cooling, imaging and hiding to enable qubit re-use experiments. 
\textbf{b,} Control infrastructure for programming quantum circuits. 
The entire waveform for all AWGs (except for rearrangement) is uploaded at the start of each experimental run. For qubit re-use experiments, the Moving, Raman AOD and Rydberg AWGs loop the same memory segment each layer. The full waveform is programmed for the Raman AWG to ensure phase continuity. For mid-circuit rearrangement, waveforms are calculated on-the-fly using a desktop computer and sent to the Rearrangement AWG operated in first-in first-out (FIFO) mode.
\textbf{c,} Level diagram of the relevant atomic transitions of $^{87}$Rb. \textbf{d,} Processor layout used for qubit re-use experiments and relevant laser beams. 
Atoms are arranged into storage, entangling and readout zones, with an additional reservoir for refilling lost atoms mid-computation. 
The 1529-nm hiding beam illuminates the storage zone to preserve coherence of active qubits during imaging in the readout zone. Parallel two-qubit gates are performed in the entangling zone with global Rydberg beams, and local detunings are optionally applied to selected gate sites using an SLM. 
The readout zone is illuminated with local beams for 1D PGC imaging and EIT cooling, as well as two counter-propagating lattice beams to form a spin-dependent potential for readout via spin-to-position conversion. 
The entire array is addressed with global Raman control for dynamical decoupling. The same Raman light is directed through a pair of crossed AODs for local single-qubit gates.
Global imaging and lambda-enhanced gray molasses cooling light are used for the initial loading. 
}
\refstepcounter{EDfig}\label{fig:ED_systemdiagram}
\end{figure*}

\begin{figure*}
\includegraphics[width=1.8\columnwidth]{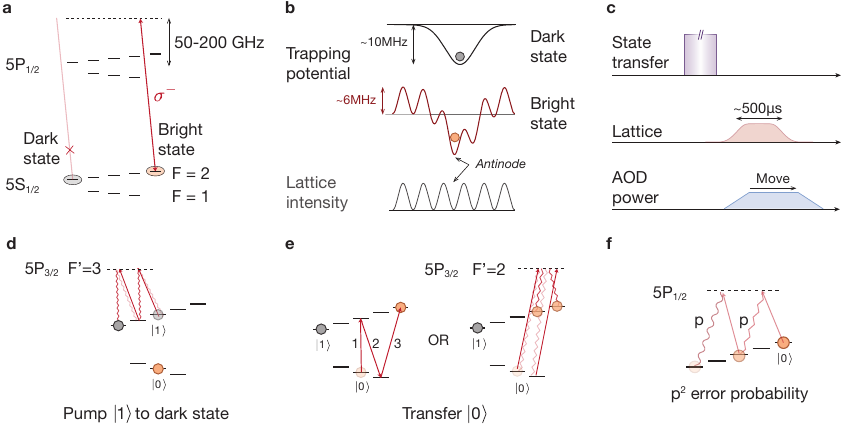}
\caption{\textbf{Spin-to-position conversion.} \textbf{a,} Level diagram showing the $^{87}$Rb hyperfine levels used to engineer a spin-selective one-dimensional optical lattice. \textbf{b,} Trapping potentials for the bright and dark states. The dark state only experiences a lightshift from the optical tweezers - allowing the atom to be moved around freely - while the bright state is additionally confined by the optical lattice potential. By using a blue-detuned lattice, atoms are trapped in intensity antinodes and so scattering of the lattice light is reduced. \textbf{c,} Schematic timeline of spin-to-position conversion. The time to transfer the clock qubit to the bright and dark states for readout is typically on the order of roughly $20\,\mu s$, but for some of our measurements is several milliseconds due to using a slow global rotation of the magnetic field (panel e). See Methods text for additional information. \textbf{d,} Transfer of qubit state $\ket{1}$ to the dark state via resonant optical pumping.
\textbf{e,} Transfer of qubit state $\ket{0}$ towards into $m_F = +1,+2$ states. This is achieved with either a coherent Raman transfer to $F=2,m_F=+2$ or incoherent pumping with $\sigma^+$-polarized repumper. Both approaches achieve the same bright state readout fidelity, but in the specific implementation we use here the Raman transfer takes several milliseconds owing to the rotation of the external magnetic field for driving $\sigma$ transitions (2 and 3). \textbf{f,} Quadratic suppression of readout error due to scattering. The bright state transfer ensures that at least two lattice-induced scattering events are required to cause a readout error, which occurs if the AOD tweezer has not yet moved away as the bright state becomes unpinned.
Scattering further causes diabatic changes in the depth of the lattice potential and may contribute to atom loss.
}
\refstepcounter{EDfig}\label{fig:ED_Lattice}
\end{figure*}

\begin{figure*}
\includegraphics[width=2\columnwidth]{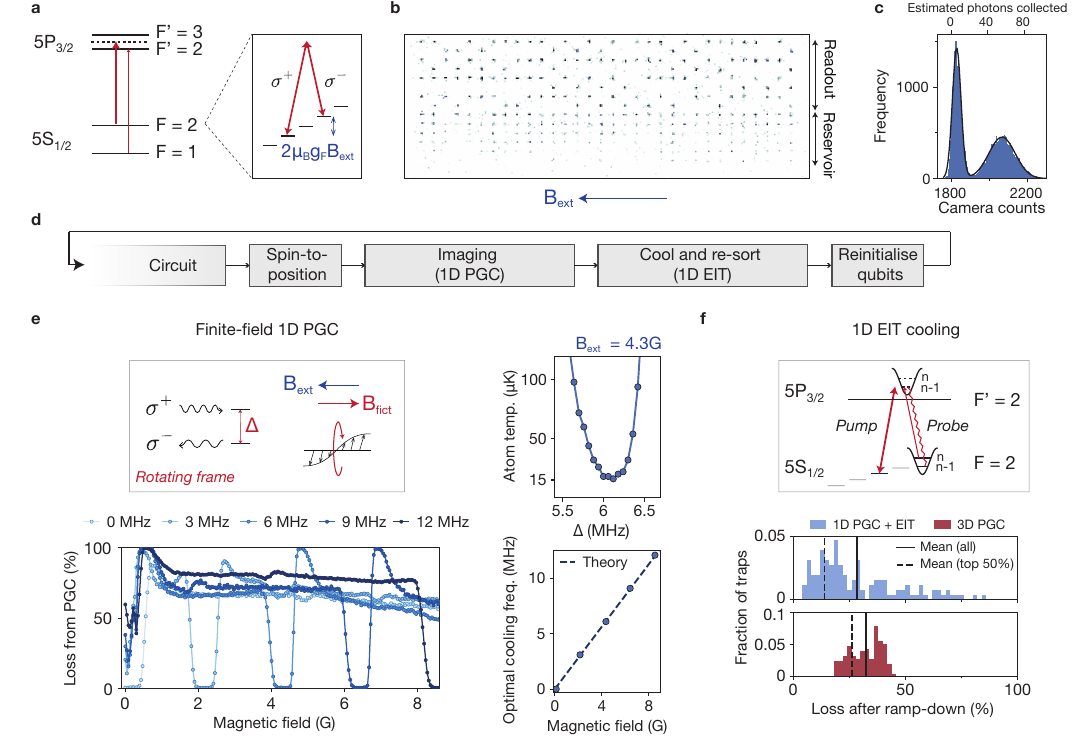}
\caption{\textbf{One-dimensional imaging and cooling in finite field.} \textbf{a,} Atomic level structure for cooling and imaging of $^{87}$Rb. The one-dimensional (1D) techniques here rely on coupling hyperfine states separated by $\delta m_F = 2$ with counterpropagating $\sigma^\pm$ beams. In a finite magnetic field $B_{ext}$, the level degeneracy is lifted by the Zeeman effect; here $\mu_B$ is the Bohr magneton and $g_F = 1/2$ the Land\'e factor. 
\textbf{b,} Single-shot local image in 8.6 G external magnetic field. Spin-to-position conversion is used in the readout zone and the reservoir is partially imaged by the tails of the imaging beams.
\textbf{c,} Example site-averaged imaging histogram in the readout zone. 
\textbf{d,} Schematic timeline of mid-circuit measurement and re-initialization used for deep circuit experiments. Due to latency bottlenecks associated with desktop-computer-based processing of images and rearrangement waveforms, we do not attempt here to optimize any speeds and simply use comfortable parameters. The circuit is 13.5-ms long. Including idle times, the spin-to-position conversion takes 4-ms, imaging takes 10-ms, cooling and re-sorting takes 13.3-ms ($\approx$7-ms latency for mid-circuit data processing), and re-initialization takes 1.1-ms. We emphasize however that these speeds can be greatly increased, as studied in the 4-ms-cycle repeated Rabi oscillations of Figure~5b.
\textbf{e,} 1D polarization gradient cooling (PGC) in finite magnetic field. Top left: interference between the two counterpropagating $\sigma^\pm$ probe beams generates a helix of linear polarization \cite{Rolston1992}. Detuning the two beams rotates the helix such that, in the rotating frame, a fictitious magnetic field appears that cancels the external field and restores the zero-field PGC cooling mechanism. Bottom left: Atom loss from 1D PGC light at different relative beam detunings, $\Delta$, in varying external magnetic field. The atoms are illuminated for 10 ms under comfortable imaging parameters. A cooling resonance is observed when $\Delta$ matches two times the Zeeman splitting (bottom right). Top right: Atom temperature around the cooling resonance in 4.3~G field, obtained from drop-recapture measurements. 
\textbf{f,} 1D electromagnetically induced transparency (EIT) cooling. Top: A strong $\sigma^+$ pump beam is combined with a weak $\sigma^-$ probe beam to drive transitions between quantum harmonic oscillator states of the optical tweezer, cooling the atom.
The EIT Fano resonance ensures heating transitions are suppressed \cite{Chow2024}.
Bottom: Ramp-down measurement of atom temperature. The SLM trap depth is adiabatically ramped down to $\sim 5\,\mu$K and held for 5-ms before being ramped back up; the atom loss probability from this process probes the temperature of all three motional axes \cite{Tuchendler2008}. Since the EIT cooling resonance is narrow and depends on trap frequency, spatial inhomogeneity in the SLM trap depths translates to inhomogeneous cooling and broadens the temperature distribution. Even so, the coldest sites after 1D PGC imaging and EIT cooling reach lower temperatures than conventional 3D PGC, highlighting the utility of these 1D techniques.}
\refstepcounter{EDfig}\label{fig:ED_ImagingAndCooling}
\end{figure*}

\begin{figure*}
\includegraphics[width=2\columnwidth]{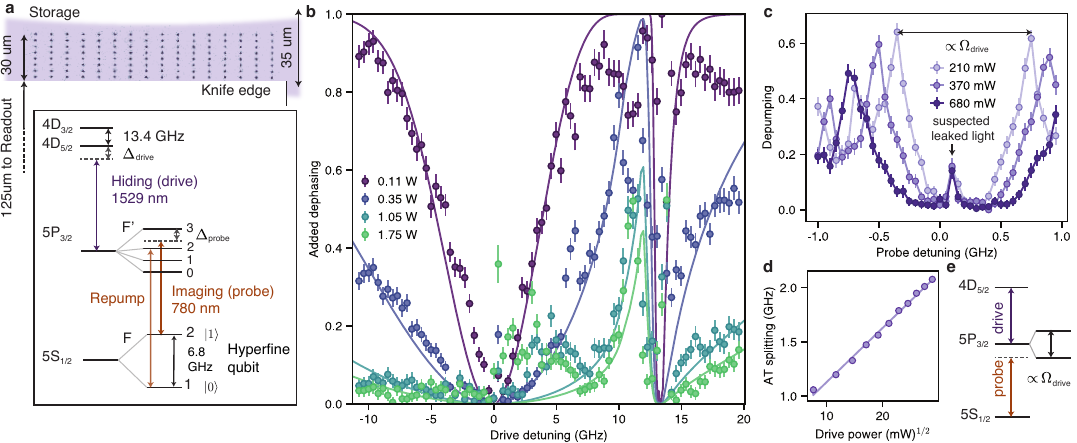}
\caption{\textbf{Atomic physics of hiding beam at 1529-nm.} \textbf{a,} Upper panel: The hiding beam is aligned to the storage zone and a knife-edge used to suppress its gaussian-tail in the readout zone. Lower panel shows the relevant atomic levels in $^{87}$Rb. The hiding beam couples the excited-to-excited state transition and imparts a 6\,GHz light-shift on the $5P_{3/2}$ state. At this detuning, the $5S_{1/2}$ ground state polarizability is approximately $2\times10^{-5}$ times smaller \cite{Hu2024}, thus maintaining coherence in the hyperfine qubit manifold while shifting the transition far off-resonance from the imaging light. \textbf{b,} Additional dephasing on atoms in the storage zone due to local imaging beams in the readout zone, for various drive powers and detunings. Here we use 20-ms of illumination and closer probe detuning than in deep circuit experiments. We plot a fit to a simple model (described in the text). For deep circuit experiments, we operate at 16\,GHz red-detuned from the bare $5P_{3/2}\rightarrow4D_{5/2}$ transition (not shown). \textbf{c,} By scanning the detuning of the probe light, we observe an Autler-Townes (AT) splitting of probe resonance when coupled to the $4D_{5/2}$-level. For this measurement we operate at a low drive power compared to typical values and the drive detuning is 500\,MHz red-detuned from resonance, such that both peaks can be experimentally measured in the limited probe detuning range. The small peak is likely due to leaked light and is not expected. \textbf{d,} The measured AT-splitting scales linearly with the Rabi frequency of the drive, or the square root of the drive power. \textbf{e,} Level diagram illustrating the coupled three-level ladder system which leads to the emergence of the Autler-Townes feature. 
}
\refstepcounter{EDfig}\label{fig:ED_1529}
\end{figure*}

\clearpage
\newpage

\begin{figure*}
\includegraphics[width=2\columnwidth]{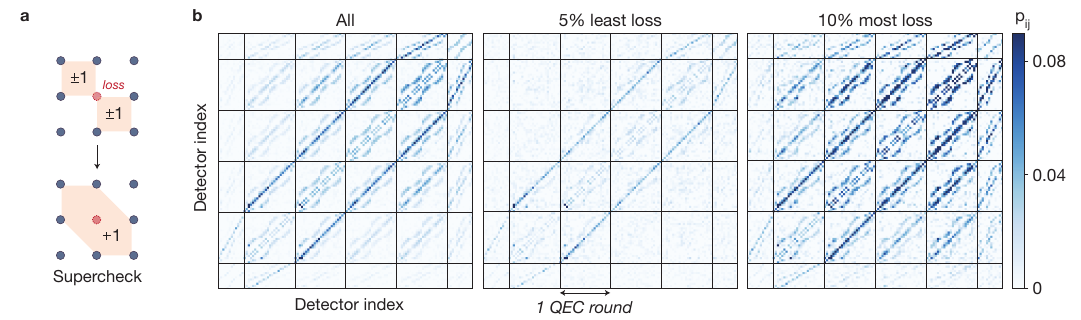}
\caption{\textbf{Characterizing effects of loss and leakage in repeated QEC.} \textbf{a,} Superchecks. A lost data qubit results in anti-commuting stabilizers and a flickering error pattern. Producting stabilizers surrounding the lost qubit recovers commuting superchecks. \textbf{b,} Detection correlation $p_{ij}$ matrix for five rounds of QEC on a $d=5$ surface code. Data qubit errors appear as space-like correlations and ancilla measurement errors as time-like correlations between adjacent layers. Leakage results in additional persistent correlations between QEC rounds. By selecting shots with the fewest lost data qubits, these correlations are suppressed, indicating that leakage is dominated by loss which can be detected. Similarly, selecting shots with the most loss enhances the correlations.}
\refstepcounter{EDfig}\label{fig:ED_SurfaceLoss}
\end{figure*}

\begin{figure*}
\includegraphics[width=2\columnwidth]{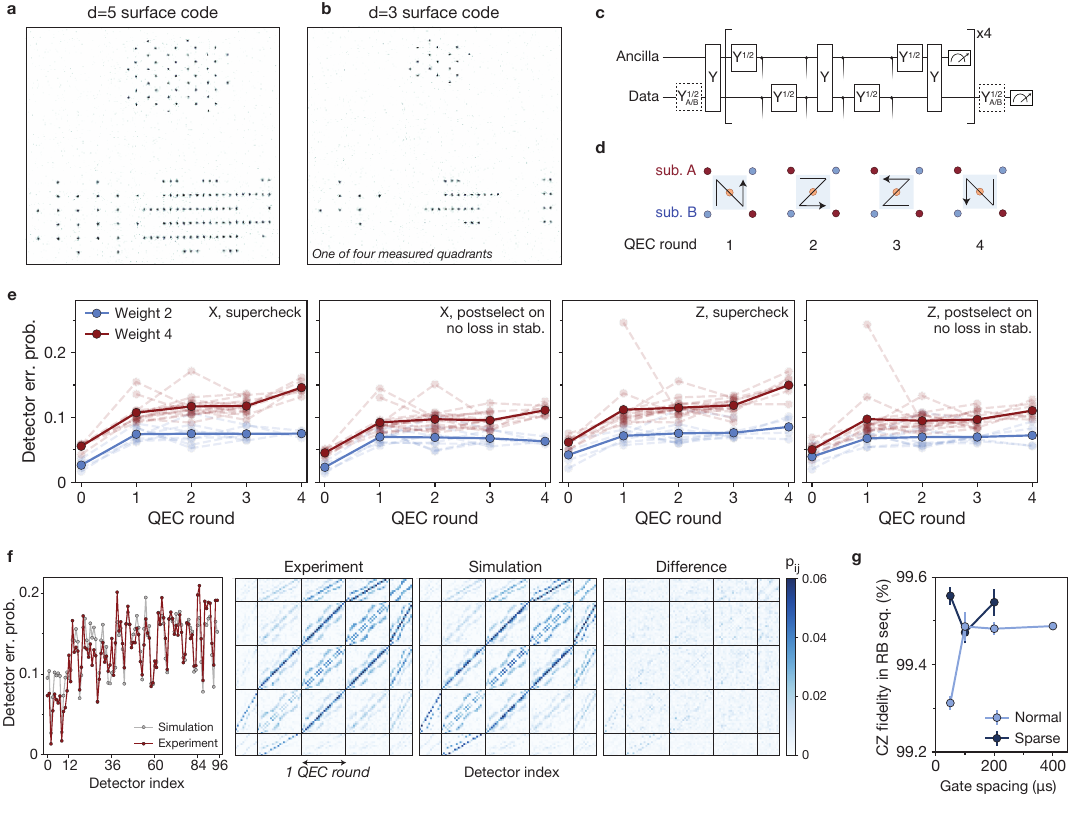}
\caption{\textbf{Additional data for repeated QEC characterization circuit.} 
\textbf{a,} Processor layout for repeated QEC on a $d=5$ surface code. The data qubits and one ancilla block are located in the entangling zone (top) and four additional ancilla blocks are in storage (bottom); one ancilla block is unused in the four-round circuit. 
\textbf{b,} Processor layout for repeated QEC on a $d=3$ code in one of four possible quadrants. 
\textbf{c,} Circuit for four rounds of QEC. For the XZZX rotated surface code \cite{BonillaAtaides2021}, Y($\pi$/2) gates are applied to one data qubit sublattice (A or B) for preparing and measuring in the X or Z basis. The equivalent circuit is obtained for stabilizer measurement of the CSS rotated surface code upon compiling Y($\pi$/2) gates. 
\textbf{d,} Stabilizer gate ordering. The same pattern is used globally in each round. 
\textbf{e,} Detector error probability for $d=5$. Faint dashed curves correspond to the 24 individual detectors (12 for the first and last rounds), and solid curves are the mean of all deterministic detectors. 
In this subfigure only, to illustrate the supercheck error distribution, we assign the error of each supercheck to all detectors from which it is composed, and plot the resulting effective detectors; the overall mean is the average of these individual detector values.
The first round has lower error due to the neighboring transversal state preparation.
The best detector over the central three rounds after loss postselection has 27\% lower error than the overall mean; one atom has anomalously high error in the Z basis. \textbf{f,} Comparison to simulation. Left: detector error probability, converting loss to qubit state 0. Right: detection correlation matrix $p_{ij}$ \cite{Quantum2023}. Both metrics show good agreement between simulation and experiment in the structure and magnitude of the errors. \textbf{g,} CZ gate fidelity measured via randomized benchmarking \cite{Evered2023a}. Infidelity due to Rydberg P states is removed by leaving sufficient time or distance between gates. Sparse corresponds to 2x larger separation between gate sites.}
\refstepcounter{EDfig}\label{fig:ED_ExtraSurface}
\end{figure*}

\begin{figure*}
\includegraphics[width=2\columnwidth]{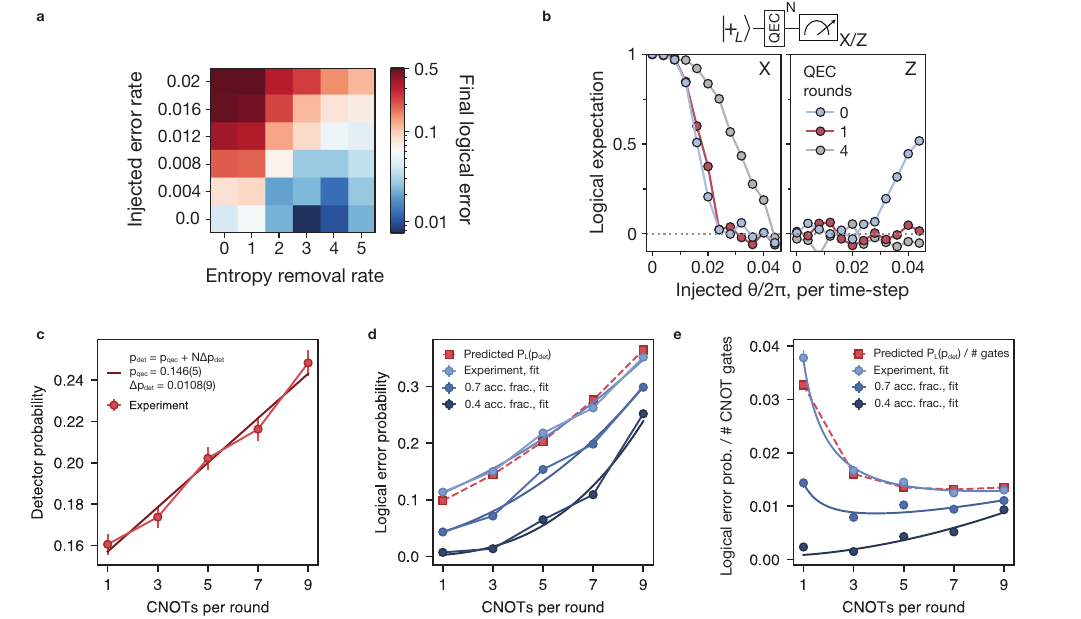}
\caption{\textbf{Exploring entropy removal during single- and two-logical-qubit operations.} \textbf{a,} Additional data for entropy removal via stabilizer measurement. As studied in Fig.~1d, the final logical error depends on the balance between the injected error rate (corresponding to the injected $\theta/2\pi$ per time-step) and the entropy removal rate (number of QEC rounds in the fixed total time window). 
For small injected error, there is an optimal number of QEC rounds (here, 3-4) since stabilizer measurement is imperfect and introduces entropy of its own.   
The plot uses MLE decoding and an acceptance fraction of 66\% to enhance salient features.
\textbf{b,} Absence of coherent logical error. Using the same error-injection protocol as in (a), the final logical state is measured in both the X and Z basis. With no QEC, the global coherent error results in a coherent logical rotation. With one round of QEC, or more, this coherent rotation vanishes. The stabilizer measurement is performed immediately after transversal state preparation, before the majority of the error is injected, such that the lack of logical rotation compared to no QEC can be attributed to the non-deterministic X(Z) stabilizer signs randomizing the response of the Z(X) logical operator to coherent error.
Here we ignore the ancillas in the MLE decoding and use a 50\% acceptance fraction. All curves are the average of both bases; for left plot, the measurement is the same basis as preparation, and for the right plot it is the orthogonal basis. \textbf{c-e} Analysis of logical gate performance of two logical qubits undergoing repeated transversal CNOTs and QEC, with the circuit studied in Fig.~3 of main text. \textbf{c,} The measured detector error probability increases linearly as a function of number of CNOTs applied in each round. \textbf{d,} Logical error probability as a function of number of CNOTs per round. Fits are to a functional form of $A\cdot(p_{\text{qec}} + N \Delta p_{\text{det}} )^3$, where $A, p_{\text{qec}},$ and $\Delta p_{\text{det}}$ are fitted parameters (blue). 
Using the fitted values of $p_{\text{qec}}$ and $\Delta p_{\text{det}}$ in \textbf{c} produces the predicted logical error probability $P_L(p_{\text{det}})$ (red). \textbf{e,} Results of $\textbf{d}$ divided by total number of CNOT gates. Logical gate fidelity $F_L(p_{\text{det}})$ is $1-P_L(p_{\text{det}}) / (3N$), where $3$ is the number of rounds and $N$ the number of gates per round.}
\refstepcounter{EDfig}\label{fig:ED_Coherent}
\end{figure*}

\begin{figure*}
\includegraphics[width=2\columnwidth]{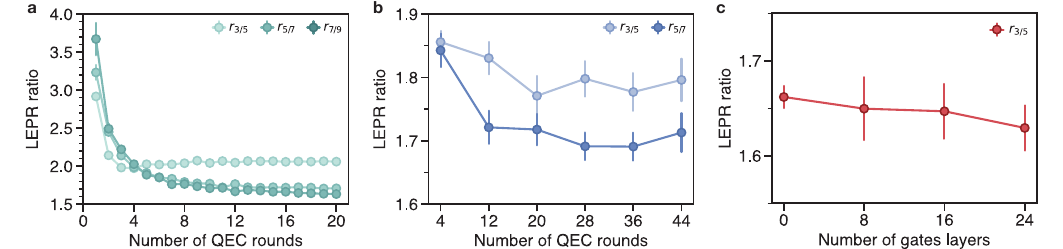}
\caption{\textbf{Theoretical characterization of logical-error-per-round ratio.} 
\textbf{a,}~Numerical simulations of rotated surface code initialized in $\ket{0_L}$ undergoing repeated QEC rounds with stabilizer measurement gate ordering in Ref.~\cite{Quantum2023}.
As a simple model, we apply a theory error channel with error probability $p=0.6\%$, where qubit resets and measurements experience uniform single-qubit depolarizing noise, CNOT gates are followed by uniform two-qubit depolarizing noise, and data qubits experience an idling single-qubit uniform depolarizing channel during ancilla qubit resets.
We plot different LEPR ratios (\mbox{$r_{d/(d+2)} = \text{LEPR}(d)/\text{LEPR}(d+2)$}) using the PyMatching decoder~\cite{Higgott2025}, observing a change of at most $17\%$ as the number of rounds is increased from four. 
The LEPR for a circuit with $N$ logical qubits and $n$ QEC rounds is defined as \mbox{$\text{LEPR} = P_{L,\text{max}}\big(1 - (1 - P_{L}/{P_{L,\text{max}}})^{1/n}\big)$}, where $P_{L,\text{max}} = 1 - \frac{1}{2^N}$ is the logical error rate of a fully mixed state.
\textbf{b,}~Numerical simulations of repeated QEC on a single $\ket{+_L}$ surface code using the same circuit as the $d=5$ surface code experiment in Fig.~3 of the main text.
We use a simplified error model where we take the experimental error model and then turn all loss rates to 0 and double the Pauli error rates for idle errors, reset, measurement, and gate errors (single and two-qubit gates) on both data and ancilla qubits.
We plot LEPR ratios for varying numbers of QEC rounds using an MLE decoder~\cite{Cain2024a}, observing changes of at most $9\%$, indicating that performance remains stable even for deeper circuits. 
\textbf{c,} 
Numerical simulations for circuits with 25 QEC rounds on four logical qubits, with interleaved transversal gate layers using the approximate experimental error model described above.
Half of the qubits, selected uniformly at random, are initialized $\ket{+_L}$, while the other half are initialized in $\ket{0_L}$.
Each gate layer consists of random pairing of transversal CNOT gates followed by logical single-qubit Pauli gates selected uniformly at random. 
After applying the random gate sequence $U$, the inverse $U^\dagger$ is applied, followed by transversal measurement in the same basis as initialization. By varying the number of gate layers interspersed between rounds, the LEPR ratio averaged over $\approx 100$ randomly sampled circuits varies by at most $2\%$.}
\refstepcounter{EDfig}\label{fig:ED_LEPR}
\end{figure*}

\begin{figure*}
\includegraphics[width=2\columnwidth]{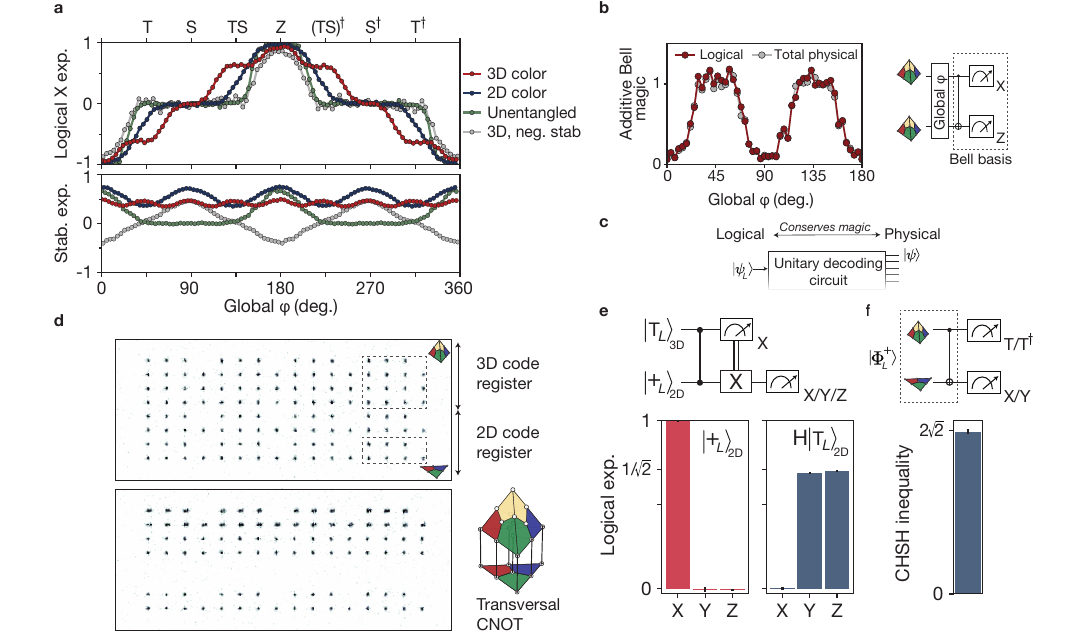}
\caption{\textbf{Universality with 3D codes.} \textbf{a,} Codes of various dimensions subject to global rotation. The same data is shown in Fig.~4a of the main text, here plotted without any normalization applied to the logical operator or stabilizer expectation values. \textbf{b,} Two-copy measurement of a [[15,1,3]] code after a global rotation \cite{Haug2023}. The additive Bell magic has a plateau at one unit of magic under a global T. The same magic is obtained by analyzing either the underlying physical 15-qubit system or the single logical qubit with error detection.
\textbf{c,} The encoded logical state is connected via a unitary Clifford decoding circuit to a product state of the encoded state and Pauli inputs on the physical qubits. The Clifford circuit conserves magic and therefore the total physical state and encoded logical state must have the same total magic. This implies that, while 15 physical T's are applied to the system, only 1 physical T is produced, which can only happen with entanglement.
\textbf{d,} Atom image of a register of 2D and 3D color codes. A transversal CNOT can be performed between the face of the [[15,1,3]] code (control) and the [[7,1,3]] (target). 
\textbf{e,} Code switching. $\ket{T_L}$ is prepared transversally on the 3D [[15,1,3]] code and then measurement and in-software feedforward teleports the logical $T$ onto a 2D [[7,1,3]] code (here also with a $H$); this illustrates the equivalence between code switching protocols and logical teleportation. Error detection is used in both plots.
\textbf{f,} Error-corrected test of Bell's inequality \cite{Bell1964}. We measure $S = \textrm{E}(X,T) + \textrm{E}(X,T^{\dagger}) + \textrm{E}(Y,T) - \textrm{E}(Y,T^{\dagger})$, where E(A,B) is the expectation value of the Steane and Reed-Muller codes in the A and B bases, respectively, and obtain S = 1.99(3)$\times \sqrt{2}$ with error detection.
}
\refstepcounter{EDfig}\label{fig:ED_Universality}
\end{figure*}

\begin{figure*}
\includegraphics[width=2\columnwidth]{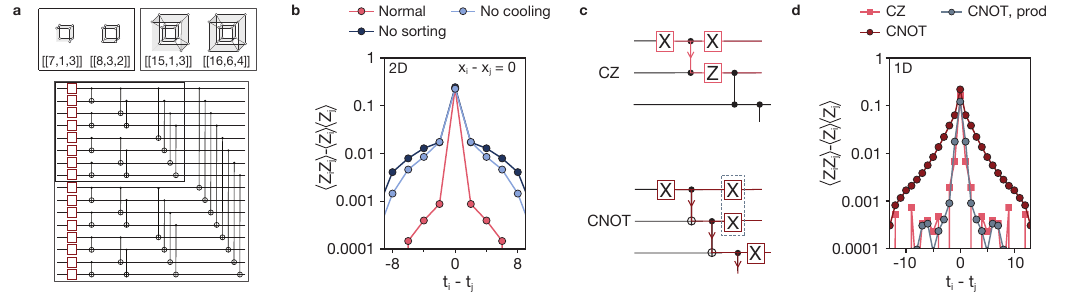}
\caption{\textbf{Entropy in deep circuits.}
\textbf{a,} General hypercube encoding. The same entangling circuit structure is combined with programmable input physical states to prepare [[7,1,3]], [[15,1,3]] and [[16,6,4]] codes (members of the family of quantum Reed-Muller codes \cite{Gong2024}). For the punctured codes, either the top or bottom qubit is removed.
\textbf{b,} Turning off entropy removal mechanisms in the 2D [[7,1,3]] cluster state circuit. In the absence of re-cooling or refilling of loss, physical stabilizer correlations persist in time. Lost atoms are assigned as qubit state 0 for this plot, and only correlations between codes constructed from the same atomic qubits in every other layer are shown.
\textbf{c,d} Circuit structure and physical error correlations. By replacing CZ gates by CNOT gates in the 1D [[7,1,3]] cluster state circuit, physical errors can propagate beyond a single layer, extending the stabilizer correlations. The product of adjacent stabilizers commutes with these propagated errors, recovering the rapid decay in physical correlations.}
\refstepcounter{EDfig}\label{fig:ED_deepcircuits}
\end{figure*}

\end{document}